\documentclass[aps,prd,twocolumn,showpacs,groupedaddress]{revtex4}  

\usepackage{graphicx}  
\usepackage{dcolumn}   
\usepackage{bm}        
\usepackage{amssymb}   
\usepackage{xspace}
\usepackage{epsfig}
\usepackage{psfrag}
\usepackage{amsfonts}
\usepackage{booktabs}
\usepackage{rotating}
\usepackage{here}

\begin{document}
\hspace{4.8in} \mbox{FERMILAB-PUB-06-386-E}

\newcommand{\etal}      {\it et al.}
\newcommand{\ttbar}     {$t\overline{t}$\xspace}
\newcommand{\elec}      {$e$\xspace}
\newcommand{\jet}       {$j$\xspace}
\newcommand{\upq}       {$u$\xspace}
\newcommand{\downq}     {$d$\xspace}
\newcommand{\strangeq}  {$s$\xspace}
\newcommand{\gluon}     {$g$\xspace}
\newcommand{\charmq}    {$c$\xspace}
\newcommand{\Jet}       {$J$\xspace}
\newcommand{\bbbar}     {$b\overline{b}$\xspace}
\newcommand{\qqbar}     {$q\overline{q}$\xspace}
\newcommand{\ttbarH}    {$t\overline{t}H$\xspace}
\newcommand{\dzero}     {D\O\xspace}
\newcommand{\met}       {\mbox{$\not\!\!E_T$}\xspace}
\newcommand{\metcal}    {\mbox{$\not\!\!E_{Tcal}$}\xspace}
\newcommand{\inpb}      {pb$^{-1}$\xspace}
\newcommand{\ppbar}     {$p\overline{p}$\xspace}
\newcommand{\qqbarprim} {$q\overline{q'}$\xspace}
\newcommand{\FERMILAB}  {\mbox{F{\scshape ermilab}}\xspace}
\newcommand{\OPAL}      {\mbox{O{\scshape pal}}\xspace}
\newcommand{\CERN}      {\mbox{C{\scshape ern}}\xspace}
\newcommand{\ATLAS}      {\mbox{A{\scshape tlas}}\xspace}
\newcommand{\herwig}    {\mbox{H\scshape{erwig}}\xspace}
\newcommand{\pythia}    {\mbox{\scshape{pythia}}\xspace}
\newcommand{\ROOT}      {\mbox{R\scshape{oot}}\xspace}
\newcommand{\jetset}    {\mbox{J\scshape{etset}}\xspace}
\newcommand{\vecbos}    {\mbox{V\scshape{ecbos}}\xspace}
\newcommand{\alpgen}    {\mbox{\scshape{alpgen}}\xspace}
\newcommand{\qq}        {\scshape{qq}}
\newcommand{\evtgen}    {\scshape{evtgen}}
\newcommand{\tauola}    {\mbox{\scshape{tauola}}\xspace}
\newcommand{\geant}     {\mbox{\scshape{geant3}}\xspace}
\newcommand{\GEANT}     {\mbox{\scshape{geant3}}\xspace}
\newcommand{\misID}     {\mbox{$\not\!\!\small{I\!D}$}}
\newcommand{\rar}       {\rightarrow}
\newcommand{\TeV}       {\mbox{\xspace$\mathrm{TeV}$}\xspace}
\newcommand{\GeV}       {\mbox{\xspace$\mathrm{GeV}$}\xspace}
\newcommand{\MeV}       {\mbox{\xspace$\mathrm{MeV}$}\xspace}
\newcommand{\pb}        {\mbox{\xspace$\mathrm{pb^{-1}}$}\xspace}
\newcommand{\fb}        {\mbox{\xspace$\mathrm{fb^{-1}}$}\xspace}
\newcommand{\SM}        {\emph {Standard Model}\xspace}
\newcommand{\NP}        {\emph {New Physics}\xspace}
\newcommand{\SMT}       {silicon vertex detector\xspace}
\newcommand{\CFT}       {central fiber detector\xspace}
\newcommand{\ETA}       {pseudorapidity\xspace}
\newcommand{\CME}       {center of mass energy\xspace}
\newcommand{\runi}      {Run~I\xspace}
\newcommand{\runii}     {Run~II\xspace}
\newcommand{\second}    {\mbox{\xspace$\mathrm{s}$}\xspace}
\newcommand{\muplus}    {$\mu$+jets\xspace}
\newcommand{\eplus}     {$e$+jets\xspace}
\newcommand{\lplus}     {$l$+jets\xspace}
\newcommand{\wplus}     {$W$+jets\xspace}
\newcommand{\zplus}     {$Z$+jets\xspace}
\newcommand{\gammaplus} {$\gamma$+jets\xspace}
\newcommand{\qcdplus}   {QCD-multijets\xspace}
\newcommand{\MET}       {$\not\!\!E_T$\xspace}
\newcommand{\wplusfour} {$W$+4jets\xspace}
\newcommand{\kapa}      {MC-to-data correction factor\xspace}
\newcommand{\kapas}     {MC-to-data correction factors\xspace}
\newcommand{\Zmumu}     {$Z\to\mu\mu$\xspace}
\newcommand{\lowmet}    {low-\mbox{$\not\!\!E_T$}-QCD\xspace}
\newcommand{\topo}      {{\em{topo}}\xspace}
\newcommand{\btag}      {{\em{btag}}\xspace}
\newcommand{\dilepton}  {$t\overline{t}_{dilepton}$\xspace}
\newcommand {\formule}[1] { \begin{eqnarray} #1 \end{eqnarray} }
\newcommand{\tbc}       {(\emph{... to be completed ...})\xspace}
\newcommand{\tbm}       {(\emph{... to be measured ...})\xspace}
\newcommand{\checkit}   {(\emph{... to be checked ...})\xspace}

\title{ Measurement of the $t\overline{t}~$ production cross section in
{\mbox{$p\bar p$}}\ collisions at {\mbox{$\sqrt{s}$ =\ 1.96\ TeV}} using secondary vertex $b$ tagging}

%
\author{                                                                      
V.M.~Abazov,$^{35}$                                                           
B.~Abbott,$^{75}$                                                             
M.~Abolins,$^{65}$                                                            
B.S.~Acharya,$^{28}$                                                          
M.~Adams,$^{51}$                                                              
T.~Adams,$^{49}$                                                              
E.~Aguilo,$^{5}$                                                              
S.H.~Ahn,$^{30}$                                                              
M.~Ahsan,$^{59}$                                                              
G.D.~Alexeev,$^{35}$
% ...
\\                                                                            
\vskip 0.30cm                                                                 
\centerline{(D\O\ Collaboration)}                                             
\vskip 0.30cm                                                                 
}                                                                             
\affiliation{                                                                 
\centerline{$^{1}$Universidad de Buenos Aires, Buenos Aires, Argentina}       
\centerline{$^{2}$LAFEX, Centro Brasileiro de Pesquisas F{\'\i}sicas,         
                  Rio de Janeiro, Brazil}                                     
\centerline{$^{3}$Universidade do Estado do Rio de Janeiro,                   
                  Rio de Janeiro, Brazil}                                     
\centerline{$^{4}$Instituto de F\'{\i}sica Te\'orica, Universidade            
                  Estadual Paulista, S\~ao Paulo, Brazil}                     
\centerline{$^{5}$University of Alberta, Edmonton, Alberta, Canada,           
                  Simon Fraser University, Burnaby, British Columbia, Canada,}
\centerline{York University, Toronto, Ontario, Canada, and                    
                  McGill University, Montreal, Quebec, Canada}                
\centerline{$^{6}$University of Science and Technology of China, Hefei,       
                  People's Republic of China}                                 
\centerline{$^{7}$Universidad de los Andes, Bogot\'{a}, Colombia}             
\centerline{$^{8}$Center for Particle Physics, Charles University,            
                  Prague, Czech Republic}                                     
\centerline{$^{9}$Czech Technical University, Prague, Czech Republic}         
\centerline{$^{10}$Center for Particle Physics, Institute of Physics,         
                   Academy of Sciences of the Czech Republic,                 
                   Prague, Czech Republic}                                    
\centerline{$^{11}$Universidad San Francisco de Quito, Quito, Ecuador}        
\centerline{$^{12}$Laboratoire de Physique Corpusculaire, IN2P3-CNRS,         
                   Universit\'e Blaise Pascal, Clermont-Ferrand, France}      
\centerline{$^{13}$Laboratoire de Physique Subatomique et de Cosmologie,      
                   IN2P3-CNRS, Universite de Grenoble 1, Grenoble, France}    
\centerline{$^{14}$CPPM, IN2P3-CNRS, Universit\'e de la M\'editerran\'ee,     
                   Marseille, France}                                         
\centerline{$^{15}$IN2P3-CNRS, Laboratoire de l'Acc\'el\'erateur              
                   Lin\'eaire, Orsay, France}                                 
\centerline{$^{16}$LPNHE, IN2P3-CNRS, Universit\'es Paris VI and VII,         
                   Paris, France}                                             
\centerline{$^{17}$DAPNIA/Service de Physique des Particules, CEA, Saclay,    
                   France}                                                    
\centerline{$^{18}$IPHC, IN2P3-CNRS, Universit\'e Louis Pasteur, Strasbourg,  
                    France, and Universit\'e de Haute Alsace,                 
                    Mulhouse, France}                                         
\centerline{$^{19}$Institut de Physique Nucl\'eaire de Lyon, IN2P3-CNRS,      
                   Universit\'e Claude Bernard, Villeurbanne, France}         
\centerline{$^{20}$III. Physikalisches Institut A, RWTH Aachen,               
                   Aachen, Germany}                                           
\centerline{$^{21}$Physikalisches Institut, Universit{\"a}t Bonn,             
                   Bonn, Germany}                                             
\centerline{$^{22}$Physikalisches Institut, Universit{\"a}t Freiburg,         
                   Freiburg, Germany}                                         
\centerline{$^{23}$Institut f{\"u}r Physik, Universit{\"a}t Mainz,            
                   Mainz, Germany}                                            
\centerline{$^{24}$Ludwig-Maximilians-Universit{\"a}t M{\"u}nchen,            
                   M{\"u}nchen, Germany}                                      
\centerline{$^{25}$Fachbereich Physik, University of Wuppertal,               
                   Wuppertal, Germany}                                        
\centerline{$^{26}$Panjab University, Chandigarh, India}                      
\centerline{$^{27}$Delhi University, Delhi, India}                            
\centerline{$^{28}$Tata Institute of Fundamental Research, Mumbai, India}     
\centerline{$^{29}$University College Dublin, Dublin, Ireland}                
\centerline{$^{30}$Korea Detector Laboratory, Korea University,               
                   Seoul, Korea}                                              
\centerline{$^{31}$SungKyunKwan University, Suwon, Korea}                     
\centerline{$^{32}$CINVESTAV, Mexico City, Mexico}                            
\centerline{$^{33}$FOM-Institute NIKHEF and University of                     
                   Amsterdam/NIKHEF, Amsterdam, The Netherlands}              
\centerline{$^{34}$Radboud University Nijmegen/NIKHEF, Nijmegen, The          
                  Netherlands}                                                
\centerline{$^{35}$Joint Institute for Nuclear Research, Dubna, Russia}       
\centerline{$^{36}$Institute for Theoretical and Experimental Physics,        
                   Moscow, Russia}                                            
\centerline{$^{37}$Moscow State University, Moscow, Russia}                   
\centerline{$^{38}$Institute for High Energy Physics, Protvino, Russia}       
\centerline{$^{39}$Petersburg Nuclear Physics Institute,                      
                   St. Petersburg, Russia}                                    
\centerline{$^{40}$Lund University, Lund, Sweden, Royal Institute of          
                   Technology and Stockholm University, Stockholm,            
                   Sweden, and}                                               
\centerline{Uppsala University, Uppsala, Sweden}                              
\centerline{$^{41}$Physik Institut der Universit{\"a}t Z{\"u}rich,            
                   Z{\"u}rich, Switzerland}                                   
\centerline{$^{42}$Lancaster University, Lancaster, United Kingdom}           
\centerline{$^{43}$Imperial College, London, United Kingdom}                  
\centerline{$^{44}$University of Manchester, Manchester, United Kingdom}      
\centerline{$^{45}$University of Arizona, Tucson, Arizona 85721, USA}         
\centerline{$^{46}$Lawrence Berkeley National Laboratory and University of    
                   California, Berkeley, California 94720, USA}               
\centerline{$^{47}$California State University, Fresno, California 93740, USA}
\centerline{$^{48}$University of California, Riverside, California 92521, USA}
\centerline{$^{49}$Florida State University, Tallahassee, Florida 32306, USA} 
\centerline{$^{50}$Fermi National Accelerator Laboratory,                     
            Batavia, Illinois 60510, USA}                                     
\centerline{$^{51}$University of Illinois at Chicago,                         
            Chicago, Illinois 60607, USA}                                     
\centerline{$^{52}$Northern Illinois University, DeKalb, Illinois 60115, USA} 
\centerline{$^{53}$Northwestern University, Evanston, Illinois 60208, USA}    
\centerline{$^{54}$Indiana University, Bloomington, Indiana 47405, USA}       
\centerline{$^{55}$University of Notre Dame, Notre Dame, Indiana 46556, USA}  
\centerline{$^{56}$Purdue University Calumet, Hammond, Indiana 46323, USA}    
\centerline{$^{57}$Iowa State University, Ames, Iowa 50011, USA}              
\centerline{$^{58}$University of Kansas, Lawrence, Kansas 66045, USA}         
\centerline{$^{59}$Kansas State University, Manhattan, Kansas 66506, USA}     
\centerline{$^{60}$Louisiana Tech University, Ruston, Louisiana 71272, USA}   
\centerline{$^{61}$University of Maryland, College Park, Maryland 20742, USA} 
\centerline{$^{62}$Boston University, Boston, Massachusetts 02215, USA}       
\centerline{$^{63}$Northeastern University, Boston, Massachusetts 02115, USA} 
\centerline{$^{64}$University of Michigan, Ann Arbor, Michigan 48109, USA}    
\centerline{$^{65}$Michigan State University,                                 
            East Lansing, Michigan 48824, USA}                                
\centerline{$^{66}$University of Mississippi,                                 
            University, Mississippi 38677, USA}                               
\centerline{$^{67}$University of Nebraska, Lincoln, Nebraska 68588, USA}      
\centerline{$^{68}$Princeton University, Princeton, New Jersey 08544, USA}    
\centerline{$^{69}$State University of New York, Buffalo, New York 14260, USA}
\centerline{$^{70}$Columbia University, New York, New York 10027, USA}        
\centerline{$^{71}$University of Rochester, Rochester, New York 14627, USA}   
\centerline{$^{72}$State University of New York,                              
            Stony Brook, New York 11794, USA}                                 
\centerline{$^{73}$Brookhaven National Laboratory, Upton, New York 11973, USA}
\centerline{$^{74}$Langston University, Langston, Oklahoma 73050, USA}        
\centerline{$^{75}$University of Oklahoma, Norman, Oklahoma 73019, USA}       
\centerline{$^{76}$Oklahoma State University, Stillwater, Oklahoma 74078, USA}
\centerline{$^{77}$Brown University, Providence, Rhode Island 02912, USA}     
\centerline{$^{78}$University of Texas, Arlington, Texas 76019, USA}          
\centerline{$^{79}$Southern Methodist University, Dallas, Texas 75275, USA}   
\centerline{$^{80}$Rice University, Houston, Texas 77005, USA}                
\centerline{$^{81}$University of Virginia, Charlottesville,                   
            Virginia 22901, USA}                                              
\centerline{$^{82}$University of Washington, Seattle, Washington 98195, USA}  
}                                                                             

\date{November 1, 2006}

\begin{abstract}
We report a new measurement of the \ttbar production cross section
in {\mbox{$p\bar p$}}
collisions at a center-of-mass energy of 1.96 TeV
using events with one charged lepton (electron or muon),
missing transverse energy,
and jets. Using
$425\;\rm pb^{-1}$ of data collected using the D0 detector at
the Fermilab Tevatron Collider, and enhancing the \ttbar content of the
sample by tagging $b$ jets with a secondary vertex tagging algorithm,
the \ttbar production cross section is measured to be:
\[
\sigma_{p\overline{p}\rightarrow t\overline{t}+X} =
6.6\pm0.9\:{\rm (stat+syst)}\:
\pm0.4\:{\rm(lum)}\:{\rm pb}.
\]
This cross section is the most precise D0 measurement to date for
\ttbar production and is
in good agreement with standard model expectations.

\end{abstract}


\maketitle

\section{Introduction}
\label{intro}
The top quark was
discovered at the Fermilab Tevatron Collider
in 1995~\cite{topdiscoveryI,topdiscoveryII}
and
completes the quark sector of the three-generation structure of the
standard model (SM). It is the heaviest known elementary particle with a mass approximately
40 times larger than that of the next heaviest quark, the bottom quark. It differs from the
other quarks not only by its much larger mass, but also by its
lifetime which is too short to build hadronic bound states.
The top quark is one of the least-studied components of the SM,
and the Tevatron, with a \CME of $\sqrt{s} = 1.96\;\rm TeV$, is at present
the only accelerator where it can be produced.
The top quark plays an important role in the
discovery of new particles, as the Higgs boson
coupling to the top quark is stronger than to all other
fermions. Understanding the signature and production rate of
top quark pairs is a crucial ingredient in the discovery
of new physics beyond the SM. In addition, it lays the ground
for measurements of top quark properties at D0.

\begin{figure}[htbp]
\centerline{\epsfxsize=3.in\epsfbox{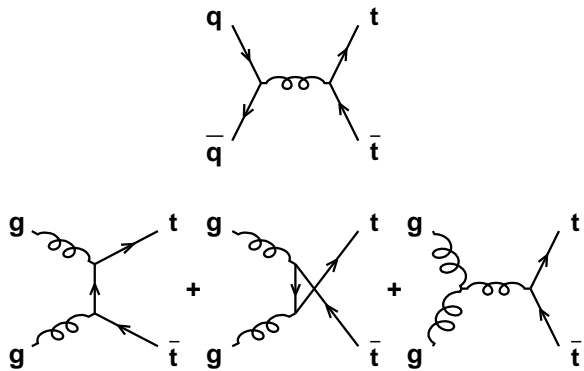}}
\caption{Leading order Feynman diagrams for the production of \ttbar pairs
at the Tevatron.}
\label{fig:topproduction}
\end{figure}

The top quark is pair-produced in $p\overline{p}$
collisions through quark-antiquark
annihilation and gluon-gluon fusion.
The Feynman diagrams of the leading order (LO) subprocesses are shown in
Fig.~\ref{fig:topproduction}.
At Tevatron energies, the $q\bar q
\rightarrow t\bar t$ process dominates, contributing 85\% of the cross
section. The $gg \rightarrow t\bar t$ process contributes the remaining 15\%.

The total top quark pair production cross section for a hard
scattering process initiated by a $p\overline{p}$ collision at the
\CME $\sqrt{s}$ is a function of the top quark mass
$m_t$ and can be expressed as
\formule
{
\sigma^{p\overline{p}\rightarrow t\overline{t}+X}(s,m_t) =
\sum\limits_{i,j=q, \overline{q}, g}
\int dx_i dx_j f_i (x_i,\mu^2) \\ \nonumber
\times \overline{f_j} (x_j,\mu^2)
\hat{\sigma}^{ij\rightarrow t\overline{t}}(\rho,m_t^2, \alpha_s(\mu^2),\mu^2).
}
The summation indices $i$ and $j$ run
over the light quarks and gluons,
$x_i$ and $x_j$ are the momentum fractions
of the partons involved in the $p\overline{p}$ collision, and
$f_i (x_i,\mu^2)$ and $\overline{f_j}(x_j,\mu^2)$ are the
parton distribution functions (PDFs) for
the proton and the antiproton, respectively.
$\hat{\sigma}^{ij\rightarrow t\overline{t}}
(\rho,m_t^2, \alpha_s(\mu^2),\mu^2)$ is the
total short distance cross section at
$\hat{s} \equiv x_i\cdot x_j\cdot s$, and is
computable as a perturbative expansion in $\alpha_s$.
The renormalization and factorization scales are chosen to be the same
parameter $\mu$, with dimensions of energy, and
$\rho \equiv \frac{4m_t^2}{\hat{s}}$.
The theoretical uncertainties on the \ttbar cross section
arise from the
choice of $\mu$ scale, PDFs, and $\alpha_s$.
For the most recent calculations of the top quark pair production cross
section, the parton-level cross sections include the full NLO matrix
elements \cite{nason}, and the
resummation of leading (LL) \cite{catani} and next-to-leading (NLL) soft
logarithms \cite{bonciani} appearing at all orders of
perturbation theory. For a top quark mass of $175\;\rm GeV$, the predicted
SM \ttbar production cross section is $6.7^{+0.7}_{-0.9}\;\rm pb^{-1}$~\cite{theoxsec}.
Deviations of the measured cross section from the theoretical
prediction could indicate effects beyond QCD perturbation theory.
Explanations might include substantial non-perturbative effects, new
production mechanisms, or additional top quark decay modes beyond the SM.
Previous measurements~\cite{CDFRun2,D0Run2,topoprd,dileptonpaper} show good agreement with the
theoretical expectation.

Within the SM, the top quark decays via the weak interaction to a $W$ boson and
a $b$ quark, with a branching fraction
$Br(t\rightarrow Wb) >$ 0.998~\cite{pdg}.
The \ttbar pair decay channels are classified as follows:
the {\em dilepton channel}, where both $W$ bosons decay leptonically
into an electron or a muon ($ee$, $\mu\mu$, $e\mu$);
the {\em \lplus channel}, where one of the $W$
bosons decays leptonically and the other hadronically ($e$+jets,
$\mu$+jets); and the {\em all-jets channel}, where both $W$ bosons decay
hadronically. A fraction of the $\tau$ leptons decays
leptonically to an electron or a muon, and two neutrinos. These events
have the same signature as events in which the $W$ boson decays
directly to an electron or a muon and are treated as part of the
signal in the \lplus channel. In addition,
dilepton events in which one of the leptons is not identified
are also treated as part of the signal in the \lplus channel.
Two $b$ quarks are present in the final state of a \ttbar event
which distinguishes it from most of the background processes.
As a consequence, identifying
the bottom flavor of the corresponding jet can be used as a selection
criteria to isolate the \ttbar signal.

This article presents a new measurement~\cite{thesisGustavo}
of the \ttbar production cross section in the
\lplus channel.
The events contain one charged lepton ($e$ or $\mu$) from a leptonic
$W$ boson decay with high transverse momentum,
missing transverse energy (\met) from the neutrino emitted in the 
$W$ boson decay,
two $b$ jets from the hadronization of the $b$ quarks, and
two non-$b$ jets ($u$, $d$, $s$, or $c$) from the hadronic $W$ decay;
additional jets are possible due to initial (ISR) and final state
radiation (FSR). $b$ jets in the event are identified by explicitly
reconstructing secondary vertices; the addition of the silicon microstrip
tracker to the upgraded detector in Run II made
this technique feasible for the first time at D0.

This paper is organized as follows: the Run II
D0 detector is described in Section~\ref{sec:detector} with special emphasis on those aspects that are
relevant to this analysis. The trigger and
event reconstruction/particle identification
techniques used to select
events that contain an electron or muon and jets are discussed in Sec.~\ref{sec:trigger} and~\ref{sec:pid}. The methods used to simulate \ttbar and
background events are explained in Sec.~\ref{sec:mc}.
A data-based method that is used to
estimate the contribution from instrumental and physics backgrounds to the
\lplus sample is presented in Sec.~\ref{sec:matrixmethod}.
The methods used to estimate the
efficiency and fake rate of the $b$ tagging algorithm are explained in Sec.~\ref{sec:SVT}.
The means for estimating all
contributions to the \lplus sample after tagging are detailed in Sec.~\ref{sec:bgr_btag}.
Finally, the description of the method used to
extract the cross section is presented in Sec.~\ref{sec:xs}.
The simulation of $W$ boson events produced in association with jets is detailed 
in Appendix~\ref{sec:wsamples}, and the handling of the statistical uncertainty on the cross section
extraction procedure is explained in Appendix~\ref{sec:stat_xs}.


\section{The D0 Detector}
\label{sec:detector}
The D0 detector \cite{d0_nim}  is a
multi-purpose apparatus designed to study \ppbar collisions at high
energies. It consists of three major subsystems.
At the core of the detector, a
magnetized tracking system precisely records the trajectories of charged
particles and measures their transverse momenta. A hermetic,
finely-grained uranium and liquid argon calorimeter measures the energies of
electromagnetic and hadronic showers.
A muon spectrometer measures the momenta of muons.

\subsection{Coordinate System}
The Cartesian coordinate system used for the D0 detector is right-handed
with the $z$ axis parallel to the direction of the protons, the $y$ axis
vertical, and the $x$ axis pointing out from the center of the accelerator ring.
A particular reformulation of the polar angle
$\theta$ is given by the \ETA
defined as
$\eta \equiv - \ln({\tan{\theta/2}})$.
In addition, the momentum vector
projected onto a plane perpendicular to the beam axis
(\textit{transverse momentum}) is defined as
$p_T = p \cdot \mathrm{sin} \theta$.
Depending on the choice of the origin of the coordinate system, the
coordinates are referred to as physics coordinates ($\phi$, $\eta$)
when the origin is
the reconstructed vertex of the interaction, or
as detector coordinates ($\phi_{\text{det}}$,
$\eta_{\text{det}}$)
when the origin is chosen to be the center of the
D0 detector.

\subsection{Luminosity Monitor\label{sec:LM}}
The Tevatron luminosity at
the D0 interaction region
is measured from the rate of inelastic
\ppbar collisions observed by the luminosity monitor (LM).
The LM consists of two arrays of
twenty-four plastic scintillator counters with photomultiplier readout.
The arrays are located in front of the forward calorimeters at $z=\pm
140\;\rm cm$  and occupy the region between the beam pipe and the forward
preshower detector. The counters are $15\;\rm cm$ long
and cover the \ETA range $2.7<|\eta_{\text{det}}|<4.4$.
The uncertainty on the luminosity
is currently estimated to be 6.1\%~\cite{lumi}.

\subsection{The Central Tracking System}
The purpose of the central tracking system~\cite{nim-tracker}
is to measure the momenta,
directions, and signs of the electric charges for charged
particles produced in a collision.
The silicon microstrip tracker (SMT) is located closest to the beam pipe
and allows for an
accurate determination of impact parameters and identification of
secondary vertices.
The length of the interaction region ($\sigma\approx 25$~cm)
led to the design of barrel modules
interspersed with disks, and assemblies of disks in the
forward and backward regions. The barrel detectors measure primarily
the $r$-$\phi$ coordinate, and the disk detectors measure $r$-$z$
as well as $r$-$\phi$.
The detector has six barrels in the central region; each barrel
has four silicon readout layers, each composed of two staggered and
overlapping sub-layers. Each
barrel is capped at high $|z|$ with a disk of twelve double-sided
wedge detectors, called an F-disk. In the far forward and backward
regions, a unit consisting of three F-disks and two large-diameter
H-disks provides tracking at high $|\eta_{\text{det}}|<3.0$.
Ionized charge is collected by
$p$ or $n$ type silicon strips of pitch
between $50$ and $150~\mu\rm m$ that are used to measure the position
of the hits.
The axial hit resolution is of the order
of $10~\mu\rm m$, the $z$ hit
resolution is $35~\mu\rm m$ for $90^\circ$ stereo and $450~\mu\rm m$ for
$2^\circ$ stereo detector modules.

Surrounding the SMT is the central fiber tracker
(CFT), which consists of $835~\mu\rm m$ diameter
scintillating fibers mounted on eight concentric support cylinders and
occupies the radial space from 20 to 52~cm from the center of the beam
pipe. The two innermost cylinders are 1.66~m long, and the outer six
cylinders are 2.52~m long. Each cylinder supports one doublet layer of
fibers oriented along the beam direction and a second doublet layer at
a stereo angle of alternating $+3^\circ$ and $-3^\circ$. In each doublet
the two
layers of fibers are offset by half a fiber width to provide improved
coverage. The CFT has a cluster resolution of
about $100~\mu\rm m$ per doublet layer.

The momenta of charged particles are determined from their
curvature in the 2~T magnetic field provided by a 2.7~m long
superconducting solenoid
magnet \cite{magnets}.
The superconducting solenoid, a two layer coil with mean
radius 60~cm, has a stored energy of 5~MJ and operates at
$10$~K. Inside the tracking
volume, the magnetic field along the trajectory of any particle
reaching the solenoid is uniform within 0.5\%. The uniformity is
achieved in the absence of a field-shaping iron return yoke by using
two grades of conductor. The superconducting solenoid coil
plus cryostat wall has a thickness of about 0.9 radiation lengths in
the central region of the detector.

Hits from both tracking detectors are combined to reconstruct tracks.
The measured momentum resolution of the tracker
can be parameterized as
$\frac{\sigma(1/p_T)}{1/p_T} =
	\sqrt{\frac{(0.003 p_T)^2}{L^4}+\frac{0.026^2}{L\sin\theta}}$,
	with the first term accounting for the  measurement
	uncertainty of the individual hits in
the tracker, and the second term for the multiple scattering.
In the expression above, $p_T$ is the particle's transverse momentum
(in GeV), and $L$ is the normalized
track bending lever arm. $L$ is equal to 1 for
tracks with $|\eta|<1.62$ and equal to $\frac{\tan\theta}{\tan\theta^{\prime}}$
otherwise. $\theta^{\prime}$ represents the angle at which the track
exits the tracker.

\subsection {The Calorimeter System}
The uranium/liquid-argon sampling calorimeters
constitute the primary system used to identify electrons, photons, and
jets. The system is
subdivided into the central calorimeter (CC) covering roughly
$|\eta_{\text{det}}|<1$ and two end calorimeters (EC) extending the coverage to
$|\eta_{\text{det}}|\approx 4$. Each calorimeter contains an electromagnetic (EM)
section closest to the interaction region, followed by fine and coarse
hadronic sections with modules that increase in size with the distance from
the interaction region. Each of the three calorimeters is located within a
cryostat that maintains the temperature at approximately $80$~K.
The EM sections use thin 3 or 4~mm plates made from
nearly pure depleted uranium. The fine hadronic sections are made from
6~mm thick uranium-niobium alloy. The coarse hadronic modules contain
relatively thick 46.5~mm plates of copper in the CC and
stainless steel in the EC.
The intercryostat region,
between the CC and the EC calorimeters, contains additional layers of
sampling, the scintillator-based intercryostat detector, to improve the energy resolution.
The CC and EC contain approximately seven and nine
interaction lengths of material respectively, ensuring
containment of nearly all particles except high $p_T$
muons and neutrinos.

The preshower detectors are designed to improve the identification
of electrons and photons and to correct for their energy losses in the
solenoid during
offline event reconstruction.
The central preshower detector (CPS) is located in the 5 cm gap between
the solenoid and
the CC, covering the region $|\eta_{\text{det}}|<$ 1.3.
The two forward preshower detectors (FPSs) are attached to the faces of the
ECs and cover the region 1.5 $<|\eta_{\text{det}}|<$ 2.5.
The relative momentum resolution for the calorimeter system
is measured in data and found to be
$\sigma(p_T)/p_T \approx 13\%$ for $50\;\rm GeV$ jets in the CC and
$\sigma(p_T)/p_T \approx 12\%$ for $50\;\rm GeV$ jets in the ECs. The energy 
resolution for electrons in the CC is 
$\sigma(E)/E \approx 15\% \sqrt{E} \oplus 4\%$.


\subsection{The Muon System}
\label{sec:muonsys}
The muon system~\cite{nim-muons}
is the outermost part of the D0 detector.
It surrounds the calorimeters and serves to identify and trigger on
muons and to provide crude measurements of momentum and charge.
It consists of a system of proportional drift tubes
(PDTs) that cover the region of $|\eta_{\text{det}}| < 1.0$ and
mini drift tubes (MDTs) that extend coverage to
$|\eta_{\text{det}}| \approx 2.0$. Scintillation counters are used for triggering and
for cosmic and beam-halo muon rejection. Toroidal magnets and
special shielding complete the muon system. Each subsystem has three
layers, with the innermost layer located between
the calorimeter and the iron of the toroid magnet. The two remaining layers are
located outside the iron. In the region directly below the CC,
only partial coverage by muon detectors is possible
to accomodate the
support structure for the detector and the readout electronics.
The average energy loss of a muon is $1.6\;\rm GeV$ in the
calorimeter and $1.7\;\rm GeV$ in the iron; the momentum
measurement is corrected for this energy loss.
The average momentum resolution for tracks that are matched to the
muon and include information from the SMT and the CFT is measured to be
$\sigma(p_T)= 0.02 \oplus 0.002 p_T$ (with $p_T$ in GeV).


\section{Triggers}
\label{sec:trigger}
The trigger system is a three-tiered pipelined system.
The first stage (Level~1) is a hardware trigger that consists
of a framework built of field programmable
gate arrays (FPGAs) which take inputs from the luminosity monitor,
calorimeter, central fiber tracker, and muon system. It makes
a decision within $4.2~\mu\rm s$ and results in a trigger accept
rate of about $2\;\rm kHz$.
In the second stage (Level~2), hardware processors associated with
specific subdetectors process information that is then used by a global processor
to determine correlations among different detectors.
Level~2 has an accept rate of 1~kHz at a maximum
dead-time of 5\% and a maximum latency of $100~\mu\rm s$.
The third stage (Level~3) uses a computing farm 
to perform a limited reconstruction of the event and make a
trigger decision using the full event information, further reducing the
rate for data recorded to tape to 50~Hz.
Throughout this analysis, the data sample
was selected at the trigger level by
requiring the presence of a lepton and a jet; however, the required
quality criteria and thresholds differ between running periods, shown in
chronological order in Table~\ref{tab:top_triggers}.

\begin{table*}[htpb]
\begin{tabular}{l|c|c|c|c}
\hline
\hline
Trigger name& $\int {\mathcal L} dt$ & Level 1 & Level 2 & Level 3 \\
 & (pb$^{-1}$) & & & \\ \hline
\multicolumn{5}{c}{\eplus channel}\\ [2.4pt]
\hline
EM15\_2JT15 & 127.8 & 1 EM tower, $E_T>10\;\rm GeV$ &
1$e$, $E_T>10\;\rm GeV$, EM fraction $>0.85$
& 1 tight $e$, $E_T>15\;\rm GeV$ \\
& & 2 jet towers, $p_T>5\;\rm GeV$ & 2 jets, $E_T>10\;\rm GeV$
 & 2 jets, $p_T>15\;\rm GeV$\\ \hline
E1\_SHT15\_2J20 & 244.0 &
1 EM tower, $E_T>11\;\rm GeV$ & None & 1 tight $e$, $E_T>15\;\rm GeV$ \\
 & & &&   2 jets, $p_T>20\;\rm GeV$\\ \hline
E1\_SHT15\_2J\_J25 & 53.7 &
1 EM tower, $E_T>11\;\rm GeV$ & 1 EM cluster, $E_T>15\;\rm GeV$
& 1 tight $e$, $E_T>15\;\rm GeV$ \\
 & & &&  2 jets, $p_T>20\;\rm GeV$ \\
 & & & & 1 jet, $p_T>25\;\rm GeV$ \\ \hline
\multicolumn{5}{c}{\muplus channel}\\ [2.4pt]
\hline
MU\_JT20\_L2M0 & 131.5 &
1~$\mu$, $|\eta_{\text{det}}|<2.0$ & 1~$\mu$, $|\eta_{\text{det}}|<2.0$  &
1 jet, $p_T>20\;\rm GeV$\\
 & &1 jet tower, $p_T>5\;\rm GeV$ &  & \\ \hline
MU\_JT25\_L2M0 & 244.0 &
1~$\mu$, $|\eta_{\text{det}}|<2.0$  & 1~$\mu$, $|\eta_{\text{det}}|<2.0$ &
1 jet, $p_T>25\;\rm GeV$\\
 & & 1 jet tower, $p_T>3\;\rm GeV$ & 1 jet, $p_T>10\;\rm GeV$ & \\ \hline
MUJ2\_JT25 & 46.2 &
1~$\mu$, $|\eta_{\text{det}}|<2.0$  & 1~$\mu$, $|\eta_{\text{det}}|<2.0$ &
1 jet, $p_T>25\;\rm GeV$\\
 &  & 1 jet tower, $p_T>5\;\rm GeV$ &
1 jet, $p_T>8\;\rm GeV$ & \\
\hline
\hline
\end{tabular}
\caption{Summary of the trigger definitions used for data collection. The
trigger
names indicate the different running periods that correspond to the
same trigger conditions. The integrated luminosity corresponding to each
running period is shown in the second column.}
\label{tab:top_triggers}
\end{table*}

Samples of events recorded with unbiased triggers are used to measure the
probability of a single object satisfying a particular trigger
requirement. Offline reconstructed objects are then identified in the
events, and the efficiency is given by the fraction of
these objects that satisfy the trigger condition
under study. Single object efficiencies are in general parameterized
as functions of the kinematic variables $p_T$, $\eta$, and $\phi$ of
the offline reconstructed objects.
The total probability for an event to satisfy a set of trigger
requirements is obtained assuming that the probability for a single
object to satisfy a specific trigger condition is independent of the
presence of other objects in the event.

The efficiency for a \ttbar event to satisfy a particular trigger
condition is measured by folding into Monte
Carlo (MC) simulated events the per-electron, per-muon, and per-jet efficiencies for
individual trigger conditions at Level~1, Level~2, and Level~3.
The total event probability $P(L1,L2,L3)$ is then calculated as the
product of the probabilities for the event to satisfy the trigger
conditions at each triggering level:
\[
P(L1,L2,L3) = P(L1) \cdot P(L2|L1) \cdot P(L3|L1,L2),
\]
where $P(L2|L1)$ and $P(L3|L1,L2)$ represent the conditional
probabilities for an event
to satisfy a set of criteria given it has
already passed the offline selection and
the requirements imposed at the previous triggering
level(s).

The overall trigger efficiency for \ttbar events corresponding to the
data samples used in this analysis
is calculated as the luminosity-weighted
average of the event probability associated with the trigger
requirements corresponding to each running period.
The systematic uncertainty on the trigger efficiency is obtained by
varying the trigger efficiency parameterizations by $\pm 1\sigma$.


\section{Event Reconstruction and Selection}
\label{sec:pid}
A collection of software algorithms
performs the offline reconstruction of each event, identifying physics 
objects (tracks, primary and secondary vertices,
electrons, photons, muons, jets and their flavor, and $\met$)
and determining their kinematic properties. Various data samples are then
selected based on the objects present in the event.
The following sections describe the
offline event reconstruction and sample selection used for this analysis.

\subsection{Tracks and Primary Vertex}
\label{sec:tracks}
Charged particles leave hits in the
central tracking system from which tracks are reconstructed.
The track reconstruction and primary vertex identification are done
in several steps:
adjacent SMT or CFT channels above a certain threshold are grouped into
clusters; sets of clusters which lie along the path of a particle are identified;
a road-based algorithm is used for track finding, followed by a Kalman 
filter~\cite{kalman} algorithm for track fitting.
The vertex search procedure~\cite{tesisAriel} consists of three steps: track clustering, track selection,
and vertex finding and fitting. First, tracks are clustered
along the $z$ coordinate, starting from the track
with the highest $p_T$ and adding tracks to the $z$-cluster if
the distance between the position along $z$ of the point of
closest approach of the track to the $z$-cluster
and the average $z$-cluster position is less than $2\;\rm cm$.
The value of this cut is optimized to effectively cluster tracks
belonging to the same interaction, while being able to
resolve multiple interactions.
Next, quality cuts are applied to the reconstructed
tracks in every $z$-cluster requiring that they have
at least 2 SMT hits, $p_T \ge 0.5\;\rm GeV$, and that they are within
three standard deviations of the nominal transverse interaction position.
Finally, for every $z$-cluster,
a tear-down vertex search algorithm fits all selected tracks to a common vertex,
excluding individual tracks from the fit until the total vertex $\chi^2$ per
degree of freedom is less than ten.
The result of the fit is a list of reconstructed vertices that contains the
hard scatter primary vertex (PV) and any additional vertices produced in
minimum bias interactions. The PV is identified from this list based on the
$p_T$ spectrum of the particles associated with each interaction.
The $\log_{10}p_T$
distribution of tracks from minimum bias processes is used to define a
probability for a track to come from a minimum bias vertex. 
The probability for a vertex to originate from a minimum bias interaction
is obtained from the probabilities for each track
and is independent of the number of tracks used in the calculation. 
The vertex with the
lowest minimum bias probability is chosen as the PV.

To ensure a high reconstruction quality for the PV, the following additional
requirements have to be satisfied:
the position along $z$ of the PV
(PV$_z$) has to be within $60\;\rm cm$ of the center of
the detector and at least three tracks have to be fitted to form
the PV. The efficiency of the PV reconstruction is about 100\% in the central
$|z|$ region, but drops quickly outside the SMT fiducial volume
($|z|<36$~cm for the barrel) due to the requirement of two SMT hits
per track forming the PV. The two
tracking detectors locate the PV with a resolution of
about $35~\mu\rm m$ along the beamline~\cite{d0_nim}.

\subsection{Electrons}
\label{sec:electrons}
Electrons are reconstructed~\cite{topoprd} using information from
the calorimeter and the central tracker.
A simple cone algorithm of radius $\Delta R=0.2$, where $\Delta R=(\Delta\phi^2+\Delta\eta^2)^{1/2}$,
clusters calorimeter cells around seeds with $E_T>$1.5 GeV.

An {\em extra-loose} electron is defined as an EM
cluster that is almost entirely contained within the EM layers
of the calorimeter, is isolated from hadronic energy depositions, and has
longitudinal and transverse shapes consistent with the expectations from
simulated electrons.
An extra-loose electron that has been spatially matched to a central track
is called a {\em loose} electron.
A loose electron is considered {\em tight} if it
passes a 7-variable likelihood test designed to distinguish between
electrons and background. The likelihood takes into
account both tracking and calorimeter information, and provides more
powerful discrimination than individual cuts on the same variables.

\subsection{Muons}
\label{sel:muons}
Muons are reconstructed using information from
the muon detector and the central tracker. Local muon
tracks are required to have hits in all three layers of the
muon system, be consistent with production in the primary collision based on
timing information from associated scintillator hits, and be located
within $| \eta_{\text{det}} | < 2.0$.
Tracks are then extended to the point of closest approach to
the beamline, and a global fit is performed considering all central tracks
within one radian in azimuthal and polar angles.
The central track with the highest $\chi^2$ probability is
assigned to the muon candidate.  The muon $p_T$, $\eta$,
and $\phi$ are
taken from the matching central track.

To reject muons from
semileptonic heavy flavor decays, the distance of closest approach
of the muon track to the PV is required to be
$<3\sigma$; in addition, the muon is required to be isolated.
Two different isolation criteria are used in this analysis~\cite{topoprd}:
the {\em loose} muon isolation criterion requires that the muon be
separated from jets, $\Delta R(\mu,{\rm jet})~>~0.5$. The {\em tight} muon
isolation criterion requires, in addition, that the muon not be surrounded by
activity in either the calorimeter or the tracker.

\subsection{Jets}
\label{sec:pid_jet}
Jets are reconstructed in the calorimeter using the
improved legacy cone algorithm~\cite{jet_algo} with radius $0.5$ and a
seed threshold of $0.5\;\rm GeV$.
A cell-selection algorithm keeps cells
with energies at least $4\sigma$ above the
average electronic noise and any adjacent cell with energy at least
$2\sigma$ above the average electronic noise ({\em T42 algorithm}).
Reconstructed jets are required to be confirmed by the independent
trigger readout, have a minimum $p_T$ of $8\;\rm GeV$, and be separated from
extra-loose electrons by $\Delta R(\text{jet},e)>0.5$.

The $p_T$ of each reconstructed jet is corrected
for calorimeter showering effects,
overlaps due to multiple interactions and event pileup, calorimeter noise,
and the energy response of the calorimeter. The calorimeter response
is measured from the $p_T$ imbalance in photon + jet events.
Jets containing a muon ($\Delta R(\mu,\text{jet})<0.5$)
are considered to originate from a semileptonic $b$ quark decay and
are corrected for
the momentum carried by the muon and the neutrino. For this correction, it
is assumed that the neutrino carries the same momentum as the muon.
The relative
uncertainty on the jet energy calibration is $\approx$ 7\% for jets with
{\mbox{20 $< p_T <$ 250 GeV}}.

\subsection{Missing $E_T$}
\label{pid_met}
The presence of a neutrino in an event is inferred
from the imbalance of the energy in the transverse plane.
This imbalance is
reconstructed from the vector sum of the transverse energies of the
cells selected by the T42 algorithm; cells of the
coarse hadronic calorimeter are only included if they are clustered
within jets. The
vector opposite to this total visible energy vector is denoted the
missing energy vector and its modulus is the raw missing transverse
energy ($\met_{\text{raw}}$).
The calorimeter missing
transverse energy ($\met_{\text{CAL}}$) is then obtained after subtracting the
electromagnetic and jet response corrections
applied to reconstructed objects in the event. Finally, the
transverse momenta of all muons present in the event
are subtracted (after correcting for the
expected energy deposition of the muon in the calorimeter) to obtain the
\met of the event.

\subsection{$b$ Jets}
\label{bid}
The secondary vertex tagging algorithm (SVT) identifies jets arising from
bottom quark hadronization ($b$ jets)
by explicitly reconstructing the decay vertex
of long-lived $b$-flavored hadrons within the jet.
The algorithm is tuned to identify $b$ jets with high efficiency,
referred to as the {\em $b$ tagging efficiency},
while keeping low the probability of
tagging a light jet (from a $u$, $d$, or $s$ quark or a gluon), referred to
as the {\em mistag rate}.
The efficiency to tag a jet
arising from
charm quark hadronization ($c$ jets)
is referred to as the
{\em $c$ tagging efficiency}.
The algorithm proceeds in
three main steps:
identification of the PV,
reconstruction of displaced secondary vertices (SVs), and the association of
SVs with calorimeter jets. The first step is described in
Sec.~\ref{sec:tracks}, the
last two steps are described below.

On average, two-thirds of the particles within a jet are electrically
charged and are therefore detected as tracks in the central tracking system.
For each track,
the distance of closest approach between the track and the beamline is
referred to as $dca$. The $z$-position of the projection of the $dca$ on the
beamline is referred to as $zdca$.
An algorithm has been developed~\cite{tesisAriel} to cluster tracks into
so-called {\em track-jets}.
Following the procedure described in Sec.~\ref{sec:tracks} 
tracks are grouped according to their
$zdca$ with respect to $z=0$.
Looping in decreasing
order of track $p_T$, tracks are added to this pre-cluster if
the difference between the track $zdca$ and the pre-cluster $z$ position
is less than $2\;\rm cm$. Next, each pre-cluster is associated with the 
vertex with the highest track multiplicity within $2\;\rm cm$ of the
center of the pre-cluster, and tracks satisfying the
following criteria are selected: $p_T>0.5\;\rm GeV$, $\geq 1$ hits in the SMT 
barrels or F-disks,
$|dca|<0.2\;\rm cm$, and $|zdca|<0.4\;\rm cm$, where
$dca$ and $zdca$ are calculated with respect to the 
reconstructed vertex associated with
the pre-cluster. Finally, for each pre-cluster,
a track-jet is formed by clustering the selected tracks
with a simple cone algorithm of radius
$\Delta R=0.5$ in $(\eta,\phi)$ space.
The procedure adds individual tracks to the jet 
cone in decreasing order of track $p_T$, and re-computes the jet
variables by adding the track 4-momentum. 
The process is repeated until no more seed tracks are left.

The secondary vertex finder is applied to every track-jet in the event with
at least two tracks.
As a first step, the algorithm loops over all tracks selecting only
      those with
      $dca$ significance $|dca/\sigma (dca)|>3.5$.
Next, the algorithm uses
a build-up method that finds two-track seed vertices by fitting
      all combinations of pairs of selected tracks within a track-jet.
      Additional tracks pointing to the seeds are attached to the vertex if
      they improve the resulting vertex $\chi^2/\text{dof}$.
      The process is
      repeated until no additional tracks can be associated with seeds. This
      procedure results in vertices that might share tracks.
The vertices found are required to satisfy the following set of
conditions: track multiplicity $\geq 2$, vertex transverse decay length
$|\vec{L}_{xy}| = |\vec{r}_{SV}-\vec{r}_{PV}|<2.6\;\rm cm$,
vertex transverse decay length significance $|L_{xy}/\sigma(L_{xy})|>7.0$,
$\chi^2_{\text{vertex}}/{\rm degrees\;of\;freedom}<10$,
and $|{\rm colinearity}|>0.9$.
The colinearity is defined as
$\vec{L}_{xy}~\cdot~\vec{p_T}^{\text{vtx}}/|\vec{L}_{xy}||\vec{p_T}^{\text{vtx}}|$,
where $\vec{p_T}^{\text{vtx}}$ is
computed as the vector sum of the momenta of all
attached tracks after the constrained fit to the secondary vertex.
The sign of the
transverse decay length is given by the sign of the
colinearity.
Secondary vertices composed of two tracks
      with opposite sign are required to be inconsistent with a
      $V^0$ hypothesis. The hypotheses tested by the algorithm include
      $K_S^0\rightarrow \pi^+\pi^-$, $\Lambda^0\rightarrow p^+\pi^-$,
      and photon conversions
      ($\gamma\rightarrow e^+e^-$). Secondary vertices are rejected if
      the invariant di-track mass is
      consistent with the tested $V^0$ mass in a mass window
      defined by $\pm 3\sigma$ of the measured $V^0$ mass resolution.

In the final step, a calorimeter jet is identified as a $b$ jet
(also called {\em tagged}) if it contains a
reconstructed SV with $L_{xy}/\sigma(L_{xy})>7.0$
within $\Delta R<0.5$. Events
containing one or more tagged jets are referred to as {\em tagged events}.

\subsection{Data Samples}
\label{sec:data_samples}
The result presented in this document is based on data recorded
using the D0 detector between August 2002 and March 2004.
Several data samples are used at various stages of the analysis and are
defined below.

The {\em \muplus preselected} sample 
is based on 422~pb$^{-1}$ of data and consists of events containing one tight
muon with $p_T>20\;\rm GeV$ and $|\eta_{\text{det}}|<2.0$
that is matched to a trigger muon, $\met>20\;\rm GeV$ separated in
$\phi$ from the muon direction, and at least one jet with
$p_T>20\;\rm GeV$ and $|\eta_{\text{det}}|<2.5$.

The {\em \eplus preselected} sample is based on 425~pb$^{-1}$ of data
and consists of events containing one tight
electron with $p_T>20\;\rm GeV$ and $|\eta_{\text{det}}|<1.1$
that is matched to a trigger electron, $\met>20\;\rm GeV$ separated in
$\phi$ from the electron direction, and at least one jet with
$p_T>20\;\rm GeV$ and $|\eta_{\text{det}}|<2.5$.

For both the \muplus and the \eplus preselected samples,
events containing a second high-$p_T$ isolated lepton are rejected to
ensure orthogonality with the dilepton analysis~\cite{dileptonpaper}.
In addition, the samples are divided into four subsamples
based on their jet multiplicity: 1, 2, or 3 jets, and 4 or more jets. In each
case, the leading jet is required to have $p_T>40\;\rm GeV$.

The preselection efficiency is measured in MC
\ttbar samples that
properly take into account tau leptons that subsequently decay
leptonically to an electron or a muon.
The efficiency measured in MC  is corrected
by data-to-MC scale factors derived from control samples where the respective
efficiency can be measured in both data and MC~\cite{topoprd}.
The quoted efficiencies include the
trigger efficiency for events that
pass the preselection, measured by folding into the MC
the per-lepton and per-jet trigger efficiencies measured in data, as
described in Sec.~\ref{sec:trigger}.
The resulting values for the preselection efficiency
for the processes $t\bar{t}\rightarrow l$+jets and
$t\bar{t}\rightarrow ll$
are summarized in Table~\ref{tab:ttbar-eff}.
\begin{table*}[!htpb]
\centering
\begin{tabular}{l|r@{$\pm$}lr@{$\pm$}lr@{$\pm$}lr@{$\pm$}l|r@{$\pm$}lr@{$\pm$}lr@{$\pm$}lr@{$\pm$}l}
\hline \hline
&\multicolumn{8}{c|}{$e$+jets}&\multicolumn{8}{c}{$\mu$+jets}\\
\hline
&\multicolumn{2}{c}{1 jet} & \multicolumn{2}{c}{2 jets} & \multicolumn{2}{c}{3 jets} & \multicolumn{2}{c|} {$\ge$ 4 jets} & \multicolumn{2}{c}{1 jet} & \multicolumn{2}{c}{2 jets} & \multicolumn{2}{c}{3 jets} & \multicolumn{2}{c} {$\ge$ 4 jets}\\
\hline
$t\bar{t}\rightarrow l$+jets & 0.79 & 0.03 & 6.02 & 0.08 & 12.99 & 0.11 &
9.01 & 0.09 & 0.52 & 0.03 & 4.67 & 0.07 & 11.66 & 0.11 & 9.20 & 0.10 \\
$t\bar{t}\rightarrow ll$ & 4.39 & 0.07 & 11.84 & 0.11 & 3.91 & 0.07 &
0.55 & 0.03 & 3.15 & 0.06 & 10.20 & 0.10 & 3.70 & 0.07 & 0.50 & 0.03 \\
\hline \hline
\end{tabular}
\caption{Summary of preselection efficiencies (\%) for \ttbar
events. Statistical uncertainties only are quoted.}
\label{tab:ttbar-eff}
\end{table*}

Systematic uncertainties in the preselection efficiencies arise from
the variation of the trigger efficiencies, the data-to-MC scale factors,
the jet energy scale and resolution,
and the jet reconstruction/identification efficiency.

In addition to the signal samples,
the following samples are selected for various studies:
The {\em muon-in-jet} sample contains two reconstructed jets
and a non-isolated muon with $\Delta R(\mu,\text{jet})<0.5$.
The {\em muon-in-jet-away-jet-tagged} sample is a subset of the muon-in-jet sample,
where the jet opposite to the one containing the muon is tagged by SVT.
The {\em EMqcd} sample contains an extra-loose electron with
$p_T>20$~GeV, at least one reconstructed jet, and \met~$\leq10$~GeV.
The {\em loose-minus-tight} sample
consists of events that pass the \eplus preselection,
except that the electron passes the loose but fails the tight selection.


\section{Event Simulation}
\label{sec:mc}
Signal and background samples are produced using the
MC event simulation methods described below. In each case,
generated events are processed through
the ~{\geant}-based~\cite{geant} ~D0
~detector simulation and reconstructed with the same program used for
collider data. Small additional corrections are applied to all
reconstructed objects to improve the
agreement between collider data and simulation.
In particular, the
momentum scales and resolutions for electrons and muons in the MC were tuned to
reproduce the corresponding leptonic
$Z$ boson invariant mass distribution observed in data, and
MC jets were smeared in energy
according to a random Gaussian distribution
to match the resolutions observed in data for the
different regions of the detector.
Overall, good agreement is observed between reconstructed objects in data
and MC.

For all MC samples, the jet flavor ($b$, $c$, or light) is determined by
matching the direction of the
reconstructed jet to the hadron flavor within the cone $\Delta R<0.5$
in $(\eta,\phi)$ space.
If there is more than one hadron found within the cone, the jet is
considered to be a $b$ jet if the cone contains at least one
$b$-flavored hadron. It is called a $c$ jet if there is at least one
$c$-flavored hadron in the cone and no $b$-flavored hadron. Light jets are required
to have no $b$ or $c$-flavored hadrons within $\Delta R <0.5$.

Production and decay of the \ttbar signal are simulated
using {\alpgen}~{\scshape 1.3}~\cite{alpgen}, which
includes the complete
$2 \to n$ partons ($2<n<6$) Born-level matrix elements, followed by
{\pythia}~{\scshape 6.2}~\cite{pythia} to simulate
the underlying event and the
hadronization. The top quark mass is set to
175~GeV. {\evtgen}~\cite{Lange:2001uf} is used to provide
the various branching fractions and lifetimes for heavy-flavor states.
The factorization and renormalization scales
for the calculation of the \ttbar process are set to
$Q = m_t$.
MC samples are generated separately for the
dilepton and \lplus signatures, according to the decay of
the $W$~bosons.
Leptons include electrons, muons, and taus, with taus
decaying inclusively using {\tauola}~\cite{Was:2004dg}.

The \wplus boson background is simulated using the same MC programs;
the factorization and renormalization scales
are set to $Q^2 = M_W^2 + \sum{(p_T^{\text{jet}})^2}$. The
events are subdivided into four disjoint samples
with 1, 2, or 3 jets, and 4 or more jets in the final state.
Details on the generation of these samples can be found in
Appendix~\ref{sec:wsamples}.

Additional samples are generated for single top quark production
(using {\sc c}omp{\sc hep}~\cite{comphep} followed by {\pythia}),
diboson production (using {\alpgen} followed by {\pythia}), and
$Z/\gamma^*\rightarrow\tau\tau$ boson production (using
{\pythia}). Since the cross sections provided by
{\alpgen} correspond to LO calculations, correction factors are
applied to scale them up to the NLO cross
sections~\cite{Campbell:1999ah}. Table~\ref{tab:xs_summary}
summarizes the generated processes with the corresponding cross
sections and NLO correction factors where applicable.
For $Z/\gamma^* \rightarrow \tau\tau$,
the cross section is quoted at NNLO and corresponds
to the mass range $60<M_Z<130\;\rm GeV$.

\begin{table}[htbp]
\begin{tabular}{l c c||c|c} \hline \hline
Process &  $\sigma$~(pb) & NLO correction & \multicolumn{2}{c}{Branching ratio} \\ \hline
 &  &  & $e$ & $\mu$ \\
\hline
{${tb \to l \nu  bb}$}  &  0.88 & --                       & 0.1259 & 0.1253 \\
{${tbq \to l \nu bbj}$} &  1.98 & --              	   & 0.1259 & 0.1253 \\
{$WW \to l \nu jj$}     &  2.04 & 1.31            	   & 0.3928 & 0.3912 \\
{$WZ \to l \nu jj$}     &  0.61 & 1.35            	   & 0.3928 & 0.3912 \\
{$WZ \to jj ll$}        &  0.18 & 1.35            	   & 0.4417 & 0.4390 \\
{$ZZ \to jjll$}         &  0.16 & 1.28            	   & 0.4417 & 0.4390 \\
{$Z/\gamma^* \to \tau\tau$} & 253 & --                     & 0.3250 & 0.3171 \\
\hline \hline
\end{tabular}
\caption{Cross sections for background processes and the
corresponding NLO correction factors, where applicable.}
\label{tab:xs_summary}
\end{table}


\section{Composition of the Preselected Samples}
\label{sec:matrixmethod}
The preselected samples are dominated by events containing a high $p_T$
isolated lepton
originating from the decay of a $W$ boson accompanied by jets. These
events are referred to as $W$-like events. The samples also include
contributions from QCD multijet events in which
a jet is misidentified as an electron (\eplus channel), or in which
a muon originating from the semileptonic decay of a heavy quark
appears isolated (\muplus channel). In addition,
substantial \met can arise from fluctuations and mismeasurements of the
jet energies. These instrumental backgrounds are referred to as the
{\em QCD multijet} background, and
their contribution is directly estimated from data, following
the {\em matrix method}.

The matrix method relies on two data sets:
a tight sample that consists of $N_t$ events that pass the preselection,
and a loose sample that consists of
$N_{\ell}$ events that pass the preselection but have the tight
lepton requirement removed, i.e., the likelihood cut for electrons
and the tight isolation requirement for muons are dropped.
The number of events with leptons originating from a $W$ boson decay is denoted
by $N^{\text{sig}}$. The number of events originating from QCD multijet
production is denoted by $N^{\text{QCD}}$. $N_{\ell}$ and $N_t$ can
be written as:
\begin{eqnarray}
\label{eq:matrix}
N_{\ell} &=& \phantom{\varepsilon_{\text{sig}}} N^{\text{sig}}+
\phantom{\varepsilon_{\text{QCD}}} N^{ \text{QCD}} \nonumber\\
N_{t}   &=& \varepsilon_{\text{sig}}   N^{\text{sig}}
+\varepsilon_{\text{QCD}} N^{ \text{QCD}}\,.
\end{eqnarray}
$\varepsilon_{\text{sig}}$ is the efficiency for a loose lepton from a $W$ boson
decay to pass the tight criteria; it is measured in \wplus MC events,
and corrected by a data-to-MC scale factor derived
from $Z\rightarrow ll $ events.
$\varepsilon_{\text{QCD}}$ is the rate at which a loose lepton
in QCD multijet events is selected as being tight; it is
measured in a low \met data sample which is dominated by
QCD multijet events.

The linear system in Eq.~\ref{eq:matrix} can be
solved for $N^{ \text{QCD}}$ and
$N^{\text{sig}}$; the
number of $W$-like events in the preselected samples is obtained as
$N_{t}^{\text{sig}}=\varepsilon_{\text{sig}}N^{\text{sig}}$, and the number of QCD multijet
events as $N_{t}^{\text{QCD}}=\varepsilon_{\text{QCD}}N^{\text{QCD}}$.
The result is summarized in Table~\ref{tab:presel_excl}.
The systematic uncertainties on the numbers of events are obtained by varying
$\varepsilon_{\text{sig}}$ and $\varepsilon_{\mathrm{QCD}}$ separately 
by one standard deviation
and adding the results of the two variations in quadrature.
As can be observed, $W$-like events
dominate the preselected samples.

\begin{table}[htbp]
\begin{center}
\begin{tabular}{lr@{$\pm$}lr@{$\pm$}lr@{$\pm$}lr@{$\pm$}l}
\hline \hline
{} & \multicolumn{2}{c} {$1~\mathrm{jet}$} & \multicolumn{2}{c}{$2~\mathrm{jets}$} & \multicolumn{2}{c}{$3~\mathrm{jets}$} & \multicolumn{2}{c}{$\geq 4~\mathrm{jets}$} \\
\hline
&\multicolumn{8}{c}{$e$+jets}\\ [2.4pt]
\hline
$N_{t}$ & \multicolumn{2}{c}{6153} & \multicolumn{2}{c} {2217} & \multicolumn{2}{c} {466}  & \multicolumn{2}{c} {119} \\
$N_{t}^{\text{sig}}$ & 5806 & 83 & 1976 & 50 & 395 & 23 & 99.8 & 11.6 \\
$N_{t}^{\text{QCD}}$ &  347 & 18 &  241 & 11 &  71 &  5 & 19.2 &  2.3 \\
\hline
&\multicolumn{8}{c}{$\mu$+jets}\\ [2.4pt]
\hline
$N_{t}$ & \multicolumn{2}{c}{6827} & \multicolumn{2}{c} {2267} & \multicolumn{2}{c} {439}  & \multicolumn{2}{c} {100} \\
$N_{t}^{\text{sig}}$ & 6607 & 85 & 2155 & 50 & 406 & 22 & 91.4 & 10.7 \\
$N_{t}^{\text{QCD}}$ &  220 & 12 &  112 & 10 &  33 &  5 &  8.6 &  2.0 \\
\hline \hline
\end{tabular}
\caption{Numbers of preselected events
and expected contributions from $W$-like and QCD multijet events
as a function of jet multiplicity. Statistical uncertainties only are quoted.}
\label{tab:presel_excl}
\end{center}
\end{table}


\section{Secondary Vertex $b$ tagging}
\label{sec:SVT}
Most of the non-\ttbar processes
found in the preselected sample do not contain heavy flavor quarks in
the final state. Requiring that one or more of the jets in the event
be tagged removes approximately
95\% of the background while keeping 60\% of the
\ttbar events. The performance of the
tagging algorithm and the methods used to determine the
corresponding efficiencies are described in this section. The efficiencies are in general parameterized
as functions of jet $p_T$ and $|\eta|$. For jets that contain a muon, the jet
$p_T$ is corrected by
subtracting the $p_T$s of the muon and the neutrino.
For this correction the neutrino is assumed to carry the same $p_T$ as the muon.
This procedure preserves the relationship
between the $p_T$ and the number of tracks in a jet which would otherwise be biased toward lower track multiplicities for jets that contain muons.

\subsection{Jet Tagging Efficiencies}
\label{sec:btag}
The probability for identifying a $b$ jet using lifetime tagging is conveniently
broken down into two components: the probability for a jet to be
taggable, called {\em taggability}, and the probability for a
taggable jet to be tagged by the SVT algorithm, called {\em tagging efficiency}.
This breakdown of the probability decouples
the tagging efficiency from issues related to detector inefficiencies,
which are absorbed into the taggability.

\subsubsection{Jet Taggability}
A calorimeter jet is considered taggable if it is matched
within $\Delta R < 0.5$ to a track-jet.
The tracks in the track-jet are required to have at least one hit in the
SMT barrel or F-disk, effectively
reducing the SMT fiducial volume to $\approx 36$ cm from the center of the
detector. Since this volume is smaller than the D0 luminous
region ($\approx 54$ cm), the taggability is expected to have a
strong dependence on the $PV_z$ of the event.
Moreover,  the relative sign between the $PV_z$ and the
jet $\eta$ must also be considered, as particular combinations of the position
of the PV
along the beam axis and the $\eta$ of the jet would enhance or reduce
the probability that a track-jet passes through the required region of the SMT.

\begin{figure}[htbp]
\begin{center}
\includegraphics[width=0.49\textwidth,clip=]{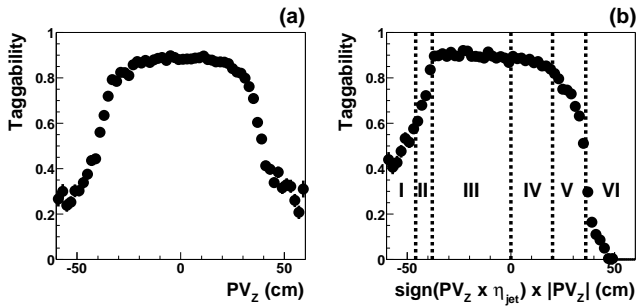}
\end{center}
\caption{Taggability vs. 
(a) $PV_z$ and (b) 
${\rm sign}(PV_z \times \eta) \times |PV_z|$ as
measured in data. The dashed lines correspond to the boundaries
between regions defined in the text.}
\label{fig:tagga_vs_pvz_data}
\end{figure}

Taggability is measured from a combined \lplus sample passing the
preselection criteria with the tight lepton requirement removed. In addition,
the $p_T$ requirement on all the jets is reduced to
$15\;\rm GeV$ to increase the statistics of the sample. No
statistically significant difference between the taggability
measured in this larger sample and directly in the \eplus and \muplus
preselected samples is observed.
Figure~\ref{fig:tagga_vs_pvz_data} shows the measured taggability as a
function of $PV_z$ and ${\rm sign}(PV_z \times \eta) \times |PV_z|$. 
The taggability decreases at the edges of the SMT barrel and this effect 
is much more pronounced when ${\rm sign}(PV_z \times \eta)>0$.
For this analysis, the taggability is parameterized as a function
of jet $p_T$ and $|\eta|$ in six bins of ${\rm sign}(PV_z \times \eta) \times |PV_z|$:
[$-60,-46$), [$-46, -38$), [$-38,0$),
[0,20), [20,36), [36,60]~(cm).
These six regions are labeled I~$-$~VI in 
Fig.~\ref{fig:tagga_vs_pvz_data}(b) and
indicated by the vertical lines. They
were chosen by taking into consideration the edge of the SMT fiducial region,
the amount of data available for the fits, and the flatness of
the taggability in each region.

\begin{figure}[htbp]
\begin{center}
\includegraphics[width=0.49\textwidth,clip=]{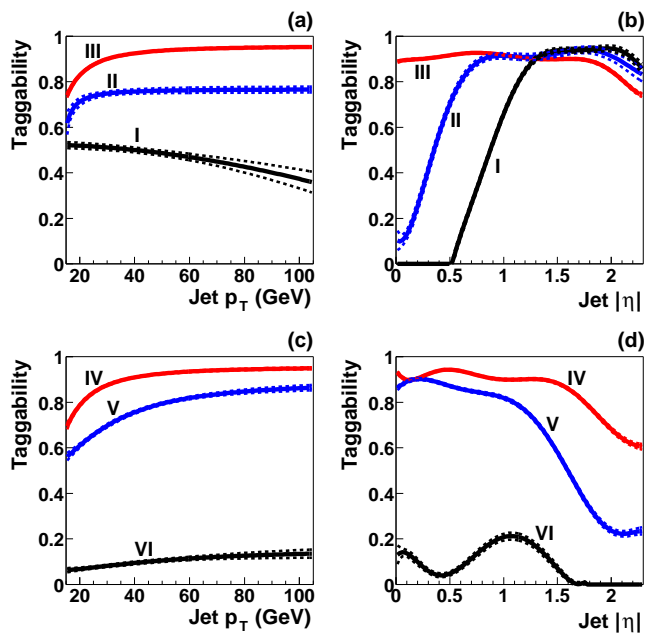}
\end{center}
\caption{Taggability vs. jet $p_T$ and $|\eta|$ for
$PV_z \times \eta < 0$ [(a) and (b) respectively] 
and $PV_z \times \eta > 0$ [(c) and (d) respectively]. 
The central value is shown with a solid line, and 
the $\pm 1 \sigma$ statistical uncertainty
is shown as dotted lines.
The labels I$-$VI correspond to the regions of
${\rm sign}(PV_z \times \eta) \times |PV_z|$ defined in
Fig.~\ref{fig:tagga_vs_pvz_data}.
\label{fig:tagga_2d}}
\end{figure}

A two-dimensional parameterization of the taggability vs. jet $p_T$ and $|\eta|$ is
derived by assuming that the dependence is factorizable, so that
$\varepsilon(p_T,\eta)=C\varepsilon(p_T)\varepsilon(\eta)$. The
normalization factor $C$ is such that
the total number of observed taggable jets equals the
number of predicted taggable jets, calculated as the
sum over all reconstructed jets weighted by their corresponding
$\varepsilon(p_T,\eta)$. Figure~\ref{fig:tagga_2d} shows $\varepsilon(p_T)$ and $\varepsilon(\eta)$ for the six regions defined
above.

The assumption that the taggability can be factorized in terms of jet
$p_T$ and $\eta$ is verified through a validation test~\cite{closure} 
that compares the
numbers of predicted and observed taggable jets as functions
of jet $p_T$, $\eta$, $PV_z$, and number of jets.
For this study, the combined $l+$jets taggability
parameterization is applied separately to the \eplus and \muplus
preselected samples as a weight for each jet.
Statistical uncertainties of the fits used to
derive the parameterizations are assigned as errors to the taggability.
Good agreement between predicted and observed distributions is observed for
all variables.

\subsubsection{Jet Flavor Dependence of Taggability}
\label{sec:taggability_hf}
\begin{figure}[htbp]
\begin{center}
\includegraphics[width=0.49\textwidth,clip=]{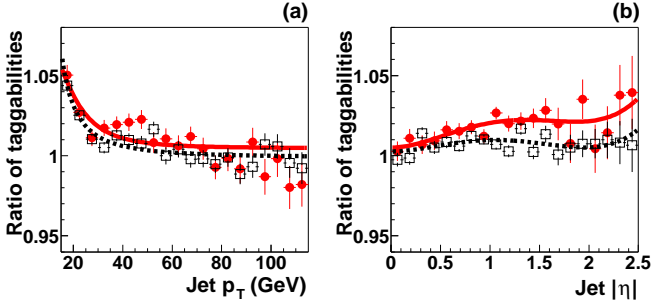}
\end{center}
\caption{Ratio of the $b$ to light (full circles) and
$c$ to light (open squares) quark jet taggability,
measured in a QCD MC sample as functions of (a) jet $p_T$ and
(b) jet $|\eta|$. 
The resulting fits used in the analysis are also shown.}
\label{fig:qcd_tag_flavor}
\end{figure}

The taggability measured in data is dominated by the predominant
light quark jet contribution to the low jet multiplicity bins.
The ratios of $b$
to light and $c$ to light taggabilities as functions of jet $p_T$
and $\eta$ are measured in a QCD multijet MC sample and shown in
Fig.~\ref{fig:qcd_tag_flavor}. The largest difference in taggability, approximately 5\%, is observed between $b$ and light quark jets in the
low $p_T$ region, corresponding to jets with low track multiplicity.
The fits to the ratios are used as flavor dependent correction
factors to the taggability.

The systematic uncertainty on the flavor dependence of the taggability
is estimated by substituting the parameterization for $b$ and $c$~quark jets
with the one determined from $Wb\bar{b}$ and $Wc\bar{c}$ MC, respectively.
The default $b$-flavor ($c$-flavor) parameterization is retained for 
the central value and the observed difference between that 
one and the $Wb\bar{b}$ ($Wc\bar{c}$) parameterization 
is taken as the systematic uncertainty.

\begin{figure}[htbp]
\begin{center}
\includegraphics[width=0.49\textwidth,clip=]{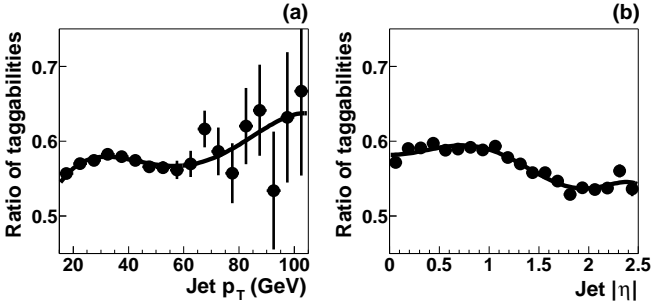}
\end{center}
\caption{Ratio of the hadronic $\tau$ to light quark jet taggability,
measured in $Z/\gamma^*\rightarrow \tau \tau$ and
$Z/\gamma^*\rightarrow q\bar{q}$ MC samples
as functions of jet $p_T$ (a) and
jet $|\eta|$ (b). The resulting fits used in the analysis are also shown.}
\label{fig:tau_tag_flavor}
\end{figure}

In comparison with light quark jets, hadronic tau lepton decays have a lower average track
multiplicity and are therefore expected to have lower taggability.
Figure~\ref{fig:tau_tag_flavor} shows the ratio of $\tau$ to light quark jet
taggability as functions of jet $p_T$ and $\eta$ as measured in
$Z/\gamma^*\rightarrow \tau \tau$ and $Z/\gamma^*\rightarrow q\bar{q}$ MC samples.
The fit to the ratio is used as a flavor dependent correction
factor to the taggability of hadronic tau decays in the
estimation of the $Z/\gamma^*\rightarrow \tau \tau$ background.

\subsection{Tagging Efficiency}
\label{sec:btagging}
The $b$ and $c$~quark jet tagging efficiencies are measured in
a \ttbar MC sample and calibrated to data using a data-to-MC
scale factor derived from a sample dominated by semileptonic $b\bar{b}$ decays.
The efficiency of tagging a light quark jet is
measured in a data sample dominated by light quark
jets and corrected for contamination of heavy flavor jets and
long-lived particles ($K_S^0$, $\Lambda^0$).
The procedures followed to determine each of the tagging efficiencies and
their corresponding uncertainties are summarized below.

\subsubsection{Semileptonic $b$ Tagging Efficiency}
\label{sec:system8}
The tagging efficiency for $b$
quarks that decay semileptonically to muons is referred to as the {\em semileptonic $b$
tagging efficiency}. It is measured in
data using a system of eight equations ({\em System8 Method}) constructed from
the total number of events in two
samples with different $b$ jet content, before and after tagging with two
$b$ tagging algorithms.
The two data samples used are
the muon-in-jet ($n$) and the
muon-in-jet-away-jet-tagged sample ($p$) (see 
Sec.~\ref{sec:data_samples} for the definition of these samples).
The two $b$ tagging algorithms are SVT and the soft lepton tagger (SLT).
The SLT algorithm requires the presence of a muon
with $\Delta R(\mu,\text{jet})<0.5$ and $p_T^{\text{rel}}>0.7$~GeV within the
jet, where
$p_T^{\text{rel}}$ refers to the muon
momentum transverse to the momentum of the jet-muon system.
The jets are divided in two categories: $b$ jets, and $c+$light ($cl$) jets,
and the following system of eight equations is written:
\begin{eqnarray}
n &=& n_b + n_{cl} \nonumber\\
p &=& p_b+ p_{cl}\nonumber\\
n^{\text{SVT}} &=&
\varepsilon_b^{\text{SVT}} n_b + \varepsilon_{cl}^{\text{SVT}} n_{cl} \nonumber\\
p^{\text{SVT}} &=&
\beta \; \varepsilon_b^{\text{SVT}} p_b + \alpha \; \varepsilon_{cl}^{\text{SVT}} p_{cl}\nonumber\\
n^{\text{SLT}} &=&
\varepsilon_b^{\text{SLT}} n_b + \varepsilon_{cl}^{\text{SLT}} n_{cl} \nonumber\\
p^{\text{SLT}} &=& \varepsilon_b^{\text{SLT}} p_b + \varepsilon_{cl}^{\text{SLT}} p_{cl} \nonumber\\
n^{\text{SVT},\text{SLT}} &=&
\kappa_b \; \varepsilon_b^{\text{SVT}} \varepsilon_b^{\text{SLT}} n_b +
\kappa_{cl} \; \varepsilon_{cl}^{\text{SVT}} \varepsilon_{cl}^{\text{SLT}} n_{cl} \nonumber\\
p^{\text{SVT},\text{SLT}} &=&
\kappa_b \; \beta \; \varepsilon_b^{\text{SVT}} \varepsilon_b^{\text{SLT}} p_b +
\kappa_{cl} \; \alpha \; \varepsilon_{cl}^{\text{SVT}} \varepsilon_{cl}^{\text{SLT}} p_{cl} \;.\nonumber
\end{eqnarray}
The terms on the left hand side represent the total number of jets
in each sample before tagging ($n$, $p$) and after tagging with
the SVT algorithm ($n^{\text{SVT}}, p^{\text{SVT}}$), the SLT algorithm
($n^{\text{SLT}}, p^{\text{SLT}}$), and both ($n^{\text{SVT,SLT}}, p^{\text{SVT,SLT}}$).
The eight unknowns on the right hand side of the
equations consist of
the number of $b$ and $c+$light jets in the two samples ($n_b$,
$n_{cl}$, $p_b$, $p_{cl}$),  and the tagging efficiencies for
$b$ and $c+$light jets for the two tagging algorithms
($\varepsilon_b^{\text{SVT}}, \varepsilon_b^{\text{SLT}},
\varepsilon_{cl}^{\text{SVT}}, \varepsilon_{cl}^{\text{SLT}}$).
The method assumes that the
efficiency for tagging a jet with both the SVT and the SLT algorithm can be
calculated as the product of the individual tagging efficiencies.
Four additional parameters are needed to solve the system of equations:
$\kappa_b$, $\kappa_{cl}$, $\alpha$, and $\beta$. The first two parameters represent
the correlation between the SVT and the SLT tagger for $b$ jets ($\kappa_b$) and
$c+$light jets ($\kappa_{cl}$), respectively. They are defined as
\begin{eqnarray*}
\kappa_b = \frac{\varepsilon_b^{\text{SVT},\text{SLT}}}
{\varepsilon_b^{\text{SVT}}\varepsilon_b^{\text{SLT}}}\,,
\end{eqnarray*}
and
\begin{eqnarray*}
\kappa_{cl} =
\frac{\varepsilon_{cl}^{\text{SVT},\text{SLT}}}{\varepsilon_{cl}^{\text{SVT}}\varepsilon_{cl}^{\text{SLT}}}\,.
\end{eqnarray*}
$\beta$ and $\alpha$ represent the ratio of the SVT tagging efficiencies
for $b$ and $c+$light jets, respectively, corresponding
to the two data samples used to solve System8.
They are defined as
\begin{eqnarray*}
\beta =\frac{\varepsilon_b^{\text{SVT}} \text{from muon-in-jet-away-jet-tagged sample}}{\varepsilon_b^{\text{SVT}} \text{from muon-in-jet sample}}\,,
\end{eqnarray*}
and
\begin{eqnarray*}
\alpha =\frac{\varepsilon_{cl}^{\text{SVT}} \text{from muon-in-jet-away-jet-tagged sample}}{\varepsilon_{cl}^{\text{SVT}} \text{from muon-in-jet sample}}\,.
\end{eqnarray*}
$\kappa_b$, $\kappa_{cl}$, and $\beta$ are measured in a MC sample mixture of
$Z/\gamma^*\rightarrow b\bar{b}\rightarrow\mu$,
$Z/\gamma^*\rightarrow c\bar{c}$, $Z/\gamma^*\rightarrow q\bar{q}$, QCD multijet, and \ttbar,
giving
$\kappa_b=0.978 \pm 0.002$, $\kappa_{cl}=0.826 \pm 0.014$, and $\beta=0.999 \pm 0.006$.
$\alpha$ is arbitrarily chosen to be $1.0\pm0.8$.

\begin{figure}[htbp]
\begin{center}
\includegraphics[width=0.49\textwidth,clip=]{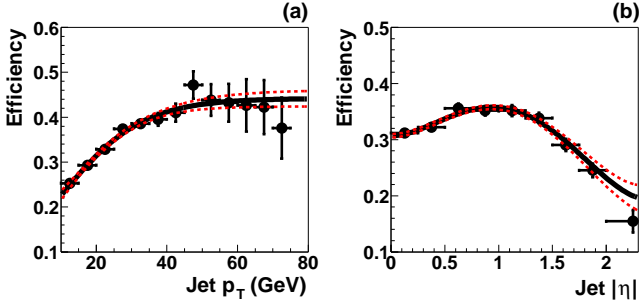}
\end{center}
\caption{Semileptonic $b$ tagging efficiency vs. jet $p_T$ (a) and jet
$|\eta|$ (b) measured in data with the System8 method.
The resulting fit is shown as a solid line,
and the $\pm 1 \sigma$ statistical uncertainty
is shown as dotted lines.}
\label{fig:svt_data_eff}
\end{figure}

The system of equations is solved for each $p_T$ and $\eta$ bin separately.
The resulting
semileptonic $b$ tagging efficiency for the SVT algorithm
is shown in Fig.~\ref{fig:svt_data_eff}.

The statistical uncertainty is given by the error on the fit to
the parameterization as functions of jet $p_T$ and $|\eta|$.
The systematic uncertainties are obtained from the
change in the semileptonic $b$ tagging efficiency resulting from the variation
on the correlation parameters $\alpha$, $\beta$, $\kappa_b$ and
$\kappa_{cl}$.
$\beta$ and $\kappa_{cl}$ are varied within the
uncertainties obtained when the distributions of
$\beta$ and $\kappa_{cl}$ as functions of jet $p_T$ are fitted to constants.
The variation of $\kappa_b$
is determined from the difference between the value of $\kappa_b$ obtained in the MC sample described above and those
obtained from
$Z/\gamma^*\rightarrow b\bar{b}$ and $t\bar{t}$ MC samples.
Another source of systematic uncertainty comes from the choice of the
$p_T^{\text{rel}}$ cut used in the SLT tagger.

\subsubsection{Measurement of the Inclusive Tagging Efficiencies}
\label{sec:incl_b-eff}
\begin{figure}[htbp]
\begin{center}
\includegraphics[width=0.49\textwidth,clip=]{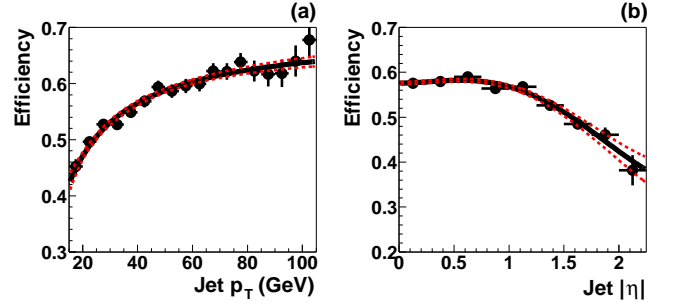}
\end{center}
\caption{Semileptonic $b$ tagging efficiency vs. (a) jet $p_T$
and (b) jet $|\eta|$ measured in a $b\bar{b}$ MC sample.
The resulting fit is shown as a solid line,
and the $\pm 1 \sigma$ statistical uncertainty
is shown as dotted lines.}
\label{fig:inclusive-eff}
\end{figure}

\begin{figure}[htbp]
\begin{center}
\includegraphics[width=0.49\textwidth,clip=]{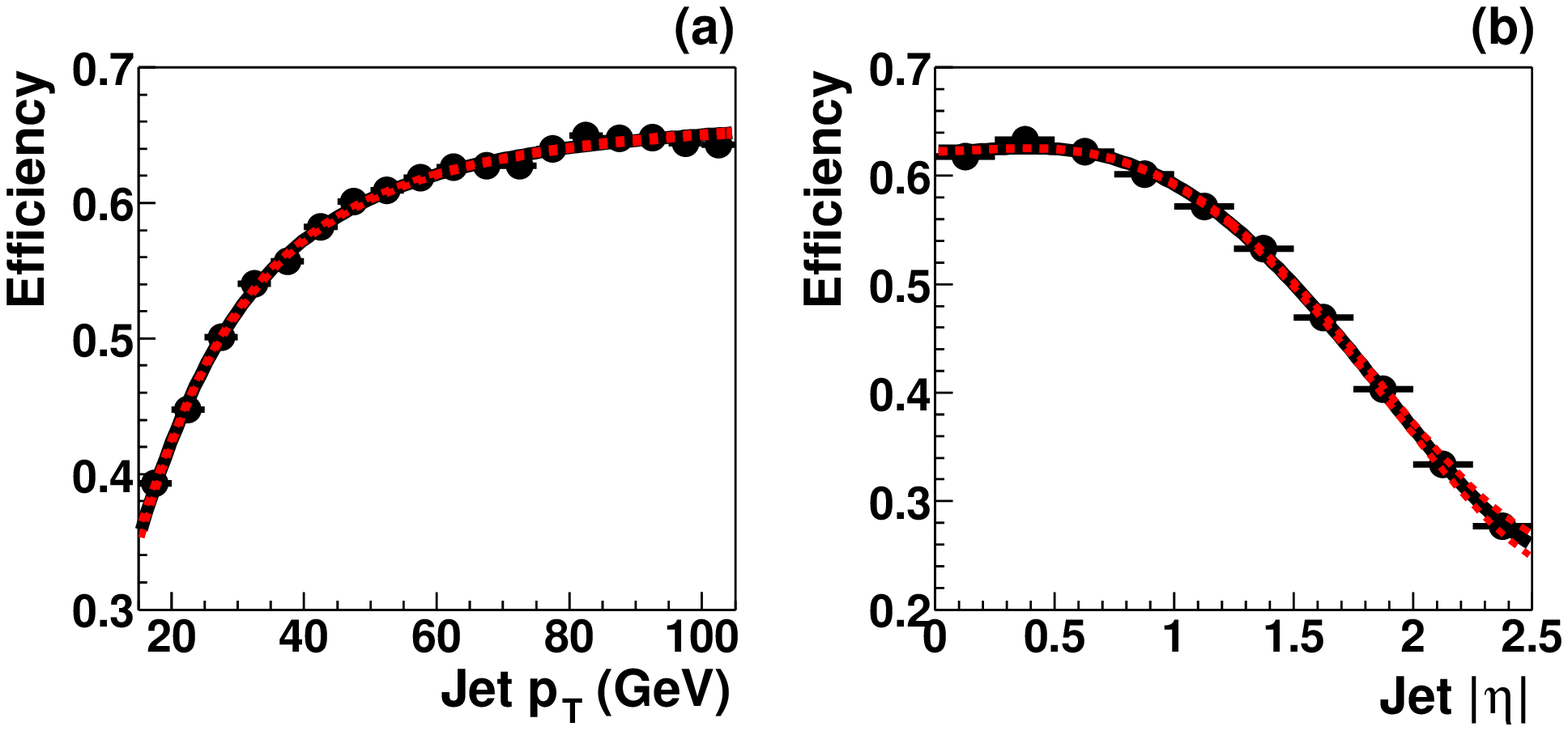}
\includegraphics[width=0.49\textwidth,clip=]{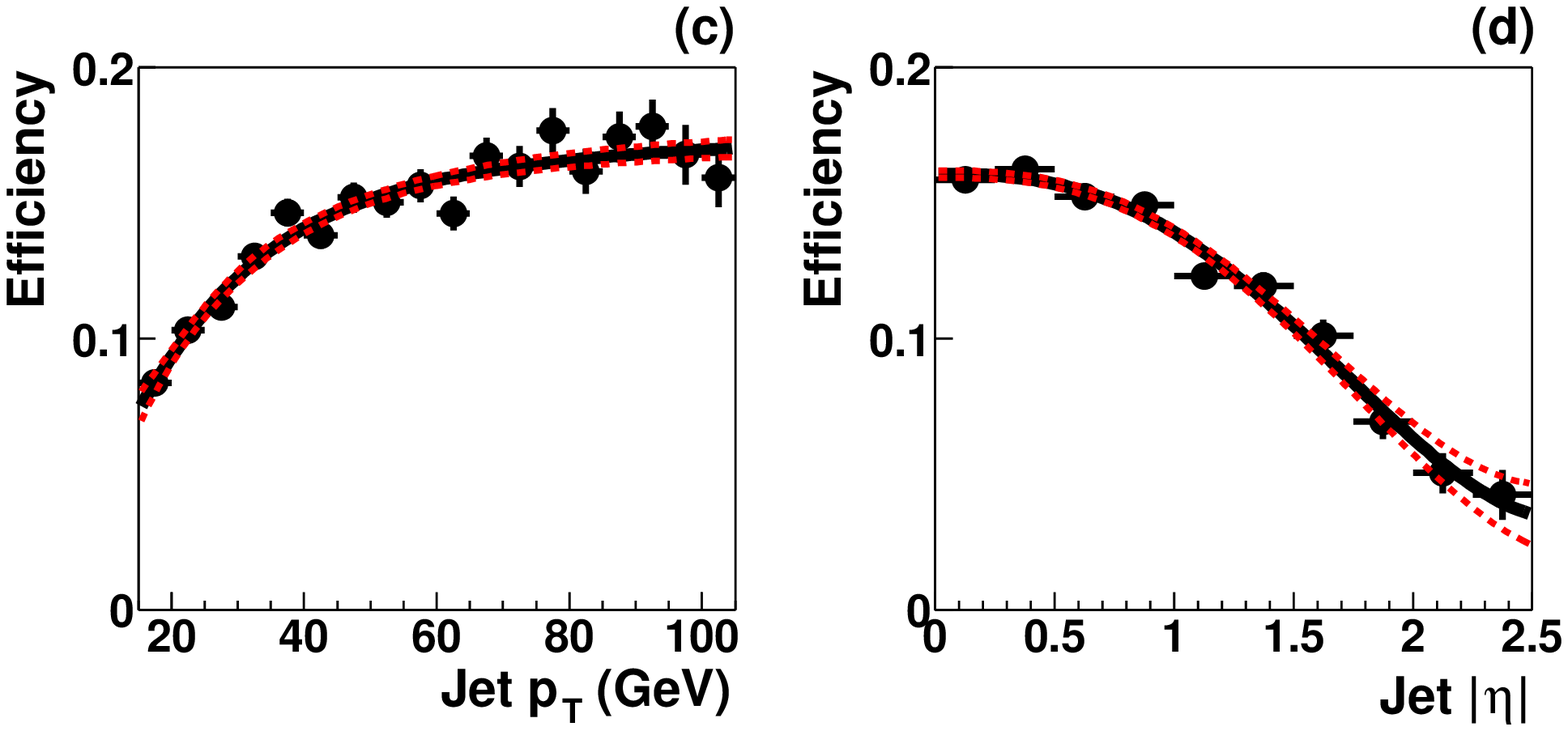}
\end{center}
\caption{
Inclusive $b$ tagging efficiency vs. (a) jet $p_T$ and (b) jet $|\eta|$ 
and inclusive $c$ tagging efficiency vs. (c) jet $p_T$ and (d) jet
$|\eta|$. The resulting fit is shown as a solid line,
and the $\pm 1 \sigma$ statistical uncertainty
is shown as dotted lines.}
\label{fig:calibrated-eff}
\end{figure}

The inclusive $b$ and $c$ tagging efficiencies are measured in a
MC \ttbar sample and calibrated
by a data-to-MC scale factor given by the ratio of the
semileptonic $b$ tagging efficiency as measured in data to the one measured in
a $b\bar{b}$ MC sample. The $b\bar{b}$ MC is chosen to determine the scale
factor because it is expected to best simulate the data samples used in
the System8 fit. With this procedure,
the topological dependence of the tagging efficiencies is taken from
the \ttbar sample, and the overall efficiency normalization is
calibrated to data.
Figure~\ref{fig:inclusive-eff} shows the semileptonic $b$ tagging efficiency
as measured in the $b\bar{b}$ MC sample.
Figure~\ref{fig:calibrated-eff} shows the inclusive $b$ and $c$ tagging
efficiencies that are used in the analysis.

The systematic uncertainty on the semileptonic $b$ tagging efficiency from
MC is taken as the difference between the 2D parameterization obtained
from $b\bar{b}$ MC and the one derived from a \ttbar MC sample. For the
inclusive $b$ and $c$ tagging efficiencies, the systematic uncertainty is
taken as the difference between the 2D parameterizations obtained
from \ttbar MC samples with two choices of $b$ fragmentation models~\cite{bFF}.
In both cases, the systematic uncertainties in each $p_T$ and $\eta$ bin
are added in quadrature to the corresponding statistical uncertainty
arising from the fit giving the default parameterization.

 A closure test~\cite{closure} 
 of the parameterized MC tagging efficiency is performed in each
case on the MC sample used to derive the default parameterization.
In addition, a validation 
is performed on a matched \wplus sample (Appendix~\ref{sec:wsamples}) that
has passed the preselection cuts. In both cases,
the predicted tags are compared with the observation as functions
of jet $p_T$, $\eta$, and jet multiplicity.
Good agreement between prediction and observation is observed in all cases.

The hadronic $\tau$ tagging efficiency is measured in a $Z/\gamma^*\rightarrow \tau \tau$ MC sample and assigned a 50\% systematic
uncertainty.
In this analysis, the hadronic $\tau$ tagging efficiency is used only
in the estimation of the $Z/\gamma^*\rightarrow \tau \tau$ background.

\subsection{Measurement of the Mistag Rate }
\label{sec:mistagging}
Mistags are defined as light flavor jets that have been tagged by
the SVT algorithm from random overlap
of tracks that are displaced from the PV due to tracking errors
or resolution effects.
Since the SVT
algorithm is symmetric in its treatment of both the impact
parameter and the decay length significance $L_{xy}/\sigma(L_{xy})$,
the mistags are expected to
occur at the same rate for {\em positive tags}
($L_{xy}/\sigma(L_{xy})>7.0$) and for
{\em negative tags} ($L_{xy}/\sigma(L_{xy})<-7.0$).
 The negative tagging rate measured in a sample dominated by light jets
 can therefore be used to extract the
 mistag rate after correcting for
the contamination of heavy flavor ($hf$) jets in the negative tags,
and the presence of long lived particles ($ll$) in the positive tags.

\begin{figure}[htbp]
\begin{center}
\includegraphics[width=0.49\textwidth,clip=]{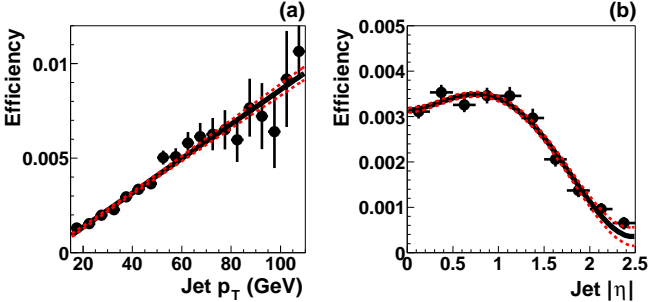}
\end{center}
\caption{Negative tagging efficiency vs. (a) jet $p_T$ and
(b) jet $|\eta|$. The resulting fit is shown as a solid line,
and the $\pm 1 \sigma$ statistical uncertainty
is shown as dotted lines.}
\label{fig:svt_ntrf}
\end{figure}

For this analysis, the negative tagging efficiency is measured
in the EMqcd data sample, which
is dominated by QCD multijet production, and parameterized as functions of
jet $p_T$ and $\eta$, as shown in Fig.~\ref{fig:svt_ntrf}.
A closure test of the parameterization is performed by comparing the
predicted rates of negative tags to the observed one in the
same sample used to derive the parameterizations.
Good agreement is observed in all distributions for jet $p_T$, $|\eta|$, and
jet multiplicity.

The parameterized negative tag rate is also applied to all taggable jets
in the preselected samples, and the prediction is compared
to the actual number of observed negative tags.
The results are summarized in Table~\ref{table:svt_ntrf_val_table} and
show good agreement between prediction and observation.

\begin{table}[htbp]
\begin{center}
\begin{tabular}{lcccc}
\hline \hline
  & 1 jet & 2 jet & 3 jet & $\ge$ 4 jet \\
\hline
\multicolumn{5}{c}{$e$+jets channel} \\ [2.4pt]
\hline
$N^{\text{pred}}$ & 24.6$\pm$5.0 & 13.4$\pm$3.7 & 3.89$\pm$1.97 & 1.54$\pm$1.24 \\
$N^{\text{obs}}$ & 22 & 16 & 5 & 4 \\
\hline
\multicolumn{5}{c}{$\mu$+jets channel} \\ [2.4pt]
\hline
$N^{\text{pred}}$ & 34.3$\pm$5.9 & 17.5$\pm$4.2 & 4.55$\pm$2.13 & 1.44$\pm$1.20 \\
$N^{\text{obs}}$ & 32 & 13 & 6 & 1 \\
\hline
\multicolumn{5}{c}{$l$+jets channel} \\ [2.4pt]
\hline
$N^{\text{pred}}$ & 58.9$\pm$7.7 & 30.9$\pm$5.6 & 8.44$\pm$2.90 & 2.98$\pm$1.73 \\
$N^{\text{obs}}$ & 54 & 29 & 11 & 5 \\
\hline \hline
\end{tabular}
\end{center}
\caption{Numbers of observed and predicted negative tags in the preselected
signal samples.}
\label{table:svt_ntrf_val_table}
\end{table}

\begin{figure}[htbp]
\begin{center}
\includegraphics[width=0.49\textwidth,clip=]{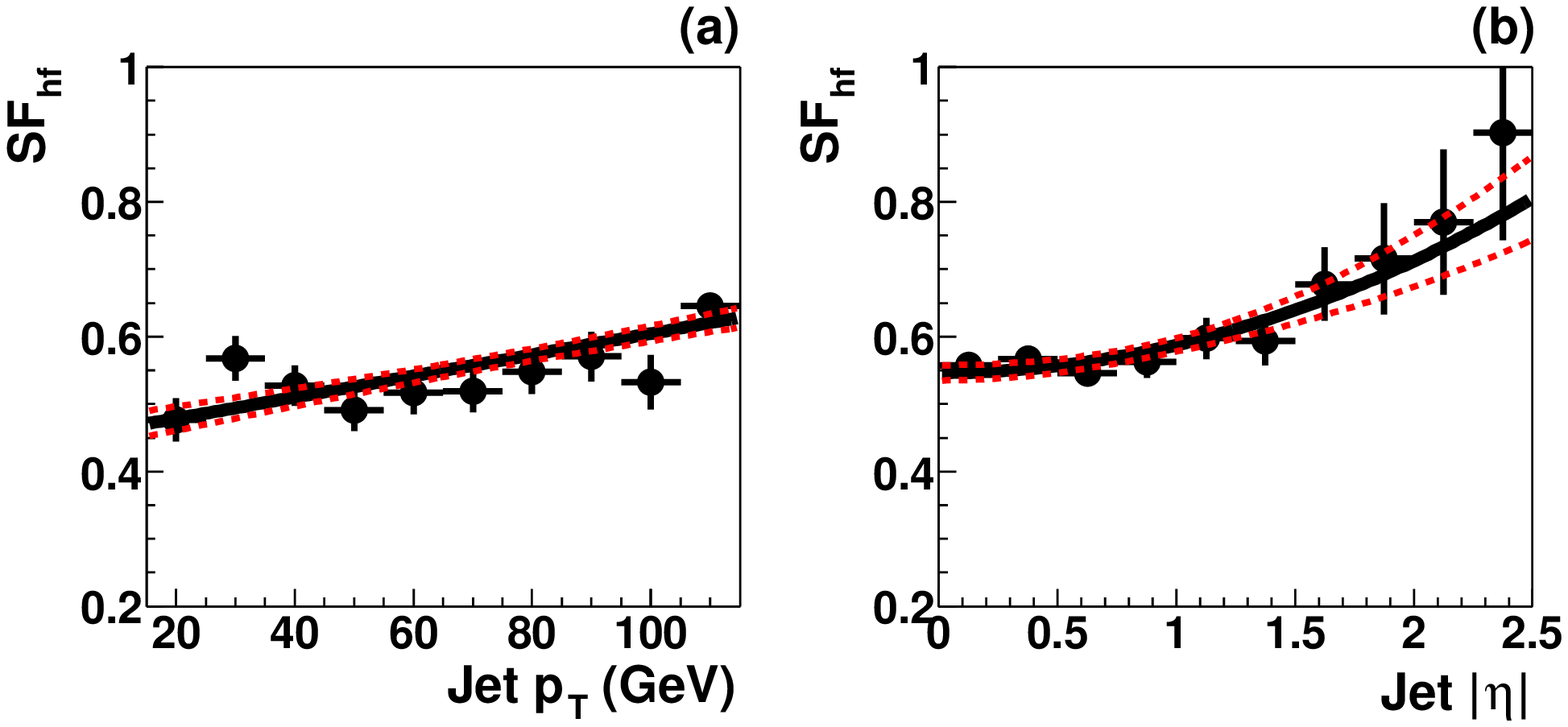}
\includegraphics[width=0.49\textwidth,clip=]{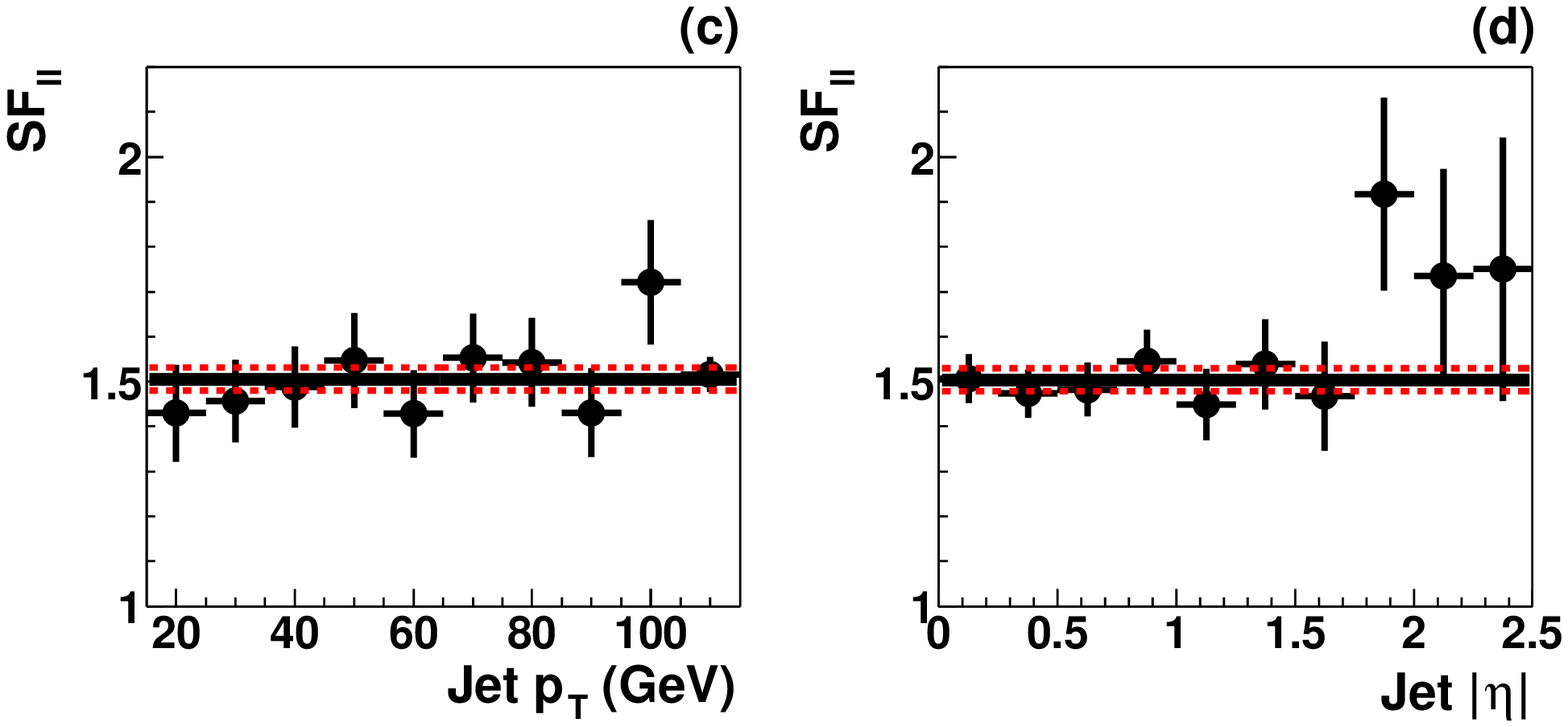}
\end{center}
\caption{Correction factors for the contribution of heavy flavor in the negative
  tag rate ($SF_{hf}$) as functions of (a) jet $p_T$ and (b) 
jet $|\eta|$ and contribution to the mistag rate from
  long lived particles ($SF_{ll}$) as functions of (c) jet $p_T$ and (d) 
jet $|\eta|$. The resulting fits are shown as solid lines,
and the $\pm 1 \sigma$ statistical uncertainties
are shown as dotted lines.}
\label{fig:svt_ntrf_SFs}
\end{figure}

To be able to use this measurement to estimate mistags from light quark jets, a correction is needed since the data sample is expected to contain
a small contribution from $b$ and $c$ jets
($\approx 2\%$ and $\approx 4\%$, respectively, as predicted by \pythia) that
have a higher negative tagging efficiency
than light quark jets.
A correction factor $SF_{hf}$ is derived from \pythia QCD multijet MC
as the ratio between the negative tagging rate for light quark jets and
the one obtained for an inclusive jet sample
\begin{eqnarray*}
SF_{hf}(p_T, \eta) = \frac{\varepsilon^{\text{light}}_{-}(p_T,
\eta)}{\varepsilon^{\text{inclusive}}_{-}(p_T, \eta)}\,.
\end{eqnarray*}
In addition, the long-lived particles
present in the EMqcd sample lead to a larger
positive than negative tagging efficiency.
A correction factor $SF_{ll}$ is derived from \pythia QCD multijet MC
as the ratio between the positive and the negative tagging rates for
light jets
\begin{eqnarray*}
SF_{ll}(p_T, \eta) = \frac{\varepsilon^{\text{light}}_{+}(p_T,
\eta)}{\varepsilon^{\text{light}}_{-}(p_T, \eta)}.
\end{eqnarray*}
Both scale factors are shown in Fig.~\ref{fig:svt_ntrf_SFs}.
Finally, the mistag rate is given by
\begin{eqnarray*}
\varepsilon^{\text{light}}_{+}(p_T, \eta)  =
\varepsilon^{\text{data}}_{-}(p_T, \eta)SF_{hf}(p_T, \eta)SF_{ll}(p_T, \eta)\,.
\end{eqnarray*}

The systematic uncertainty on the mistag rate is
determined by coherently varying by 20\%
the $b$ and $c$ fractions in the \pythia QCD multijet MC sample used to
measure $SF_{hf}$ and $SF_{ll}$. The resulting
systematic uncertainty in each $p_T$ and $\eta$ bin
is added in quadrature to the corresponding statistical uncertainty
arising from the fit giving the default parameterization
for $\varepsilon^{\text{data}}_{-}$,
$SF_{hf}$, and $SF_{ll}$.


\subsection{Event Tagging Probability}
\label{sec:evttagprob}

The probability for a jet of a given
flavor $\alpha$ ($b$, $c$, or light quark jet) to be tagged is obtained as
the product of the taggability and the calibrated tagging efficiency
\begin{eqnarray*}
{\mathcal P}_{\alpha}(p_T, \eta) = P^{\text{taggab}}_{\alpha}(p_T, \eta)\varepsilon_{\alpha}(p_T, \eta).
\end{eqnarray*}

The probability for a given MC event to contain at least one SVT-tagged jet
is given by the complement of the probability
      that none of the jets is tagged:
\begin{eqnarray*}
P^{\text{tag}}_{\text{event}}(\ge 1\;\mathrm{tag})
= 1-P^{\text{tag}}_{\text{event}}(0\;\mathrm{tag})\,,
\end{eqnarray*}
with
\begin{eqnarray*}
P^{\text{tag}}_{\text{event}}(0\;\mathrm{tag})
= \prod_{j=1}^{N_{\text{jets}}}[1-{\mathcal P}_{\alpha_{j}}(p_{T_{j}}, \eta_{j})]\,.
\end{eqnarray*}
      The probabilities for a given MC event to have exactly one or to
      have two or more SVT tagged jets are
      given by
\begin{eqnarray*}
P^{\text{tag}}_{\text{event}}(1\;\mathrm{tag}) = \sum_{j=1}^{N_{\text{jets}}}{{\mathcal
P}_{\alpha_{j}}(p_{T_{j}}, \eta_{j})} \prod_{i \ne j}[1-{\mathcal
P}_{\alpha_{i}}(p_{T_{i}}, \eta_{i})]\,,
\end{eqnarray*}
      and
\begin{eqnarray*}
P^{\text{tag}}_{\text{event}}(\ge 2\;\mathrm{tag}) = P^{\text{tag}}_{\text{event}}(\ge 1\;\mathrm{tag}) - P^{\text{tag}}_{\text{event}}(1\;\mathrm{tag})\,,
\end{eqnarray*}
      respectively.
$P^{\text{tag}}_{\text{event}}(1\;\mathrm{tag})$ and $P^{\text{tag}}_{\text{event}}(\ge 2\;\mathrm{tag})$
are referred to as single and double tagging probabilities, respectively.

The average event
tagging probability for a certain process
$P^{\text{tag}}_{\text{process}}$ is calculated by averaging the
per-event SVT tagging probability over a sample of events
for the process under consideration.
The probability for an event to satisfy the trigger conditions is
included in the calculation, as the trigger
can distort the jet $p_T$ and $\eta$ spectra,
particularly for the low jet multiplicity bins.

The trigger-corrected average event tagging probability is measured
for MC \ttbar events that pass the preselection and originated from
the processes
$t\bar{t}\rightarrow l$+jets and $t\bar{t}\rightarrow ll$;
the results are summarized in Table~\ref{tab:ttbar}.

\begin{table*}[!htpb]
\centering
\begin{tabular}{l|r@{$\pm$}lr@{$\pm$}lr@{$\pm$}lr@{$\pm$}l|r@{$\pm$}lr@{$\pm$}lr@{$\pm$}lr@{$\pm$}l}
\hline \hline
&\multicolumn{8}{c|}{$e$+jets}&\multicolumn{8}{c}{$\mu$+jets}\\
\hline
 & \multicolumn{2}{c} {$1~\mathrm{jet}$} & \multicolumn{2}{c}{$2~\mathrm{jets}$} & \multicolumn{2}{c}{$3~\mathrm{jets}$} & \multicolumn{2}{c}{$\geq 4~\mathrm{jets}$} & \multicolumn{2}{c} {$1~\mathrm{jet}$} & \multicolumn{2}{c}{$2~\mathrm{jets}$} & \multicolumn{2}{c}{$3~\mathrm{jets}$} & \multicolumn{2}{c}{$\geq 4~\mathrm{jets}$}\\
\hline
\multicolumn{17}{c}{\ttbar single tag probabilities (\%)} \\ [2.4pt] \hline
$t\bar{t}\rightarrow l$+jets & 26.6 & 0.7 & 38.7 & 0.2 & 43.3 & 0.1 & 44.7 & 0.1 & 26.2 & 0.9 & 37.8 & 0.2 & 42.7 & 0.1 & 44.1 & 0.1 \\
$t\bar{t}\rightarrow ll$ & 38.8 & 0.2 & 44.7 & 0.1 & 44.9 & 0.2 & 44.6 & 0.5 & 38.4 & 0.3 & 44.0 & 0.1 & 44.5 & 0.2 & 44.1 & 0.5 \\
\hline
\multicolumn{17}{c}{\ttbar double tag probabilities (\%)} \\ [2.4pt] \hline
$t\bar{t}\rightarrow l$+jets & \multicolumn{2}{c}{} & 4.93 & 0.10 & 11.5 & 0.1 & 15.4 & 0.1 & \multicolumn{2}{c}{}&
5.06 & 0.11 & 11.5 & 0.1 & 15.2 & 0.1 \\
$t\bar{t}\rightarrow ll$ &\multicolumn{2}{c}{} & 12.4 & 0.1 & 13.6 & 0.1 & 14.1 & 0.4 & \multicolumn{2}{c}{} & 12.1 & 0.1 & 13.6 & 0.1 & 13.5 & 0.4 \\
\hline \hline
\end{tabular}
\caption{Summary of the average event tagging probabilities
(\%) for \ttbar events that pass the preselection and originate from
the processes $t\bar{t}\rightarrow l$+jets and $t\bar{t}\rightarrow ll$.
Statistical uncertainties only are quoted.}
\label{tab:ttbar}
\end{table*}


\section{Composition of the Tagged Sample}
\label{sec:bgr_btag}
The main background to the tagged \lplus sample is
heavy flavor production in association with a $W$ boson.
Additional contributions arise from direct QCD heavy flavor production,
other low rate electroweak processes
(single top, diboson, and $Z/\gamma^*\to \tau \tau$ production),
as well as mistags of light quark jets. The methods used to
estimate the contribution from these
background processes are described below.

\subsection{Evaluation of the \wplus Background}
Available MC
generators are able to perform matrix element calculations for
\wplus events with high jet multiplicities only at leading order. As a result,
the overall normalization of the calculations suffers from large theoretical
uncertainties, although the relative contributions of the different
processes are well described. In this analysis, the overall normalization
of the \wplus contribution is obtained directly from collider data,
and only the relative contributions of different processes are taken from MC.
The contribution of \wplus events to the
tagged sample is then estimated by multiplying the
number of \wplus events of each type in the preselected sample by the
SVT efficiency
corresponding to the type of process under consideration, as described below.

The overall normalization of the $W$-like background in the preselected
sample before tagging ($N^{\text{sig}}_{t}$) is obtained directly from
collider data as described in Sec.~\ref{sec:matrixmethod}.
$N^{\text{sig}}_{t}$ consists mostly of \wplus background events, with contributions
from \ttbar and other low rate electroweak processes.
Thus, the number of \wplus events in the preselected sample can be
calculated as
\begin{eqnarray*}
N^{\text{presel}}_{W+\text{jets}} & = & N^{\text{sig}}_{t}-N_{t\bar{t}\rightarrow
l+\text{jets}}^{\text{presel}}-N_{t\bar{t}\rightarrow ll}^{\text{presel}} \\ \nonumber
& - & \sum_{\text{bkg}\;i}{N}_{\text{bkg}\;i}^{\text{presel}}\,,
\end{eqnarray*}
where $i$ loops over the electroweak backgrounds.
It is important to note
that $N_{t\bar{t}\rightarrow l+\text{jets}}^{\text{presel}}$ and
$N_{t\bar{t}\rightarrow ll}^{\text{presel}}$
are allowed to
float during the extraction of the \ttbar cross section, adjusting the
\wplus contribution accordingly.

The predicted number of \wplus events in the tagged sample
is obtained by multiplying the estimated number of preselected
\wplus events by the corresponding average event tagging probability
$P^{\text{tag}}_{W+\text{jets}}$:
\begin{eqnarray*}
N^{\text{tag}}_{W+\text{jets}} = N^{\text{presel}}_{W+\text{jets}} P^{\text{tag}}_{W+\text{jets}}\,.
\end{eqnarray*}
$P^{\text{tag}}_{W+\text{jets}}$ is obtained
by adding the tagging probabilities for the different flavor configurations
considered, weighted by their fractions within a given jet multiplicity bin
\begin{eqnarray*}
P^{\text{tag}}_{W+\text{jets}} = \sum_{\Phi_n} F_{\Phi_n}\,
P^{\text{tag}}_{\Phi_n}.
\end{eqnarray*}
$F_{\Phi_n}$ gives the fraction of events that pass the preselection for
each flavor configuration $\Phi$ per jet multiplicity
bin $n$. It is determined by:
\begin{eqnarray*}
F_{\Phi,n} = \frac{\sigma^{\mathrm{eff}}_{\Phi,n}}{\sum_{\Phi} \sigma^{\mathrm{eff}}_{\Phi,n}}\,,
\end{eqnarray*}
where
$\sigma^{\mathrm{eff}}_{\Phi,n} \equiv \sigma_{\Phi,n} \cdot \varepsilon^{\text{presel},\text{match}}_{\Phi,n}$
is the effective cross section, obtained by multiplying the
theoretical cross section $\sigma_{\Phi,n}$ from {\alpgen} by the
preselection and matching efficiency
$\varepsilon^{\text{presel},\text{match}}_{\Phi,n}$ for each flavor configuration and
jet multiplicity.
The flavor configurations considered in the analysis were
identified according to the ad hoc
matching prescription discussed in Appendix~\ref{sec:wsamples} and are
summarized in Table~\ref{tab:flav_configs}.
$P^{\text{tag}}_{\Phi_n}$ is the corresponding average event tagging
probability, as defined in Sec.~\ref{sec:evttagprob}.
The resulting
event tagging probabilities for each $W$+jets flavor subprocess are
summarized in Table~\ref{tab:svt_tag_prob}.

\begin{table*}[htbp]
\begin{center}
\begin{tabular}{lr@{$\pm$}lr@{$\pm$}lr@{$\pm$}lr@{$\pm$}l}
\hline \hline
{Contribution} & \multicolumn{2}{c}{$W$+1 jet} & \multicolumn{2}{c}{$W$+2 jets} & \multicolumn{2}{c}{$W$+3 jets}
& \multicolumn{2}{c}{$W$+$\ge$4 jets} \\
\hline
$Wb\bar{b}$             & \multicolumn{2}{c}{} & $1.23$&$ 0.08$ & $2.05$&$ 0.21$ & $2.84$&$0.16$ \\
$Wc\bar{c}$             & \multicolumn{2}{c}{} & $1.69$&$ 0.12$ & $2.94$&$ 0.37$ & $4.44$&$0.29$ \\
$W(b\bar{b})$           &  $0.86$&$0.03$ & $1.46$&$0.09$ & $2.03$&$0.15$ & $2.99$&$0.24$ \\
$W(c\bar{c})$           &  $1.23$&$0.05$ & $2.26$&$0.15$ & $3.08$&$0.24$ & $5.06$&$0.54$ \\
$Wc$                    &  $4.41$&$0.18$ & $6.25$&$0.43$ & $4.93$&$0.48$ & $4.30$&$0.23$ \\
$W$+light               &  $93.5$&$ 0.2$ & $87.1$&$0.7$  & $85.0$&$1.1$  & $80.4$&$0.7$  \\
\hline \hline
\end{tabular}
\caption{Fractions (\%) of different \wplus flavor subprocesses contributing
to each jet multiplicity bin when ad hoc matching and preselection
are required. Statistical uncertainties only are quoted.}
\label{tab:flav_configs}
\end{center}
\end{table*}

The choice of cone size used for the ad hoc matching procedure
contributes to the systematic  uncertainty.
To estimate this
effect, the cone size is varied from the
default value of $\Delta R=0.5$ to $\Delta R=0.7$,
and the difference,
centered on the default value,
is assigned as the systematic uncertainty on the fractions.
This results in a relative uncertainty 
of 2\% for the $Wc$ fractions and 
5\% for the $Wb\bar{b}$, $W(b\bar{b})$, $Wc\bar{c}$,
and $W(c\bar{c})$
fractions, in all jet multiplicities 
(refer to Appendix~\ref{sec:wsamples} for a definition of these samples).  
In addition, the \wplus fractions are also derived from
limited-statistics MC samples
where matrix element partons are matched to particle
jets following the MLM matching scheme~\cite{Mangano}. The difference between
the fractions obtained from these samples and the ones derived from samples
matched with the ad hoc method
is less than 20\% for the region of interest
(events with three or more jets), and does
not depend on the choice of matching parameters.
An additional
20\% systematic uncertainty is assigned to the \wplus fractions based
on this study.

The fractions calculated with both matching procedures are
obtained from MC samples based on LO calculations.
Several studies~\cite{mcfm,Campbell} of $W$+2 jets processes have
established that the ratio of $Wb\bar{b}$ to $Wjj$ cross
sections at NLO is higher
by a factor $K=1.05\pm 0.07$ compared to the LO prediction.
The systematic uncertainty on the $K$-factor arises
from the residual dependence on the factorization scale and from
the uncertainty on the PDFs, which is obtained using the
20 eigenvector pairs for the {\scshape CTEQ6M} PDFs~\cite{CTEQ6}.
This $K$-factor is applied to correct
the ad hoc fractions of $Wb\bar{b}$, $W(b\bar{b})$, $Wc\bar{c}$,
and $W(c\bar{c})$, while for the $Wc$ fraction, the LO prediction is used.
The fraction of $W+$light jets is adjusted to ensure that the
sum of all fractions equals 1.

Additional systematic uncertainties associated with the $W$ boson modeling
arise from the choice of parton distribution functions, factorization scale,
and heavy quark mass. The systematic uncertainty arising from
each of these factors on the
$W$+jets fractions is calculated from the relative change in the
{\alpgen} cross section, properly taking correlations into account.
The PDF uncertainty is calculated using the 20 eigenvector pairs from
{\scshape CTEQ6M}; the factorization scale uncertainty is calculated
by varying the scale to two times and one-half of the default value;
the heavy quark mass uncertainty is calculated by
varying by $\pm 0.3$ GeV~\cite{pdg} 
the heavy quark masses with respect to
their default values ($m_b=4.75$ GeV and $m_c=1.55$ GeV).

An alternative method of obtaining the event tagging probability
for $W$+light jets
is to apply the light tagging efficiency parameterization directly to the
preselected signal sample. Under the assumption that the preselected sample
is dominated by $W$+light jets events, this method has the advantage of
taking the
kinematic information directly from the data. The event tagging
probabilities obtained with this alternative method are also shown in Table~\ref{tab:svt_tag_prob} and are in good agreement with those obtained from MC.

The expected number of
$W$+jets events for each flavor subprocess as a function of jet multiplicity are summarized in Tables~\ref{tab:svt_summary_table_1}
and~\ref{tab:svt_summary_table_2}
for single and double tagged events, respectively.

\begin{table*}[htbp]
\centering
\begin{tabular}{l|r@{$\pm$}lr@{$\pm$}lr@{$\pm$}lr@{$\pm$}l|r@{$\pm$}lr@{$\pm$}lr@{$\pm$}lr@{$\pm$}l}
\hline \hline
& \multicolumn{8}{c|}{$e$+jets}&\multicolumn{8}{c}{$\mu$+jets} \\
\hline
 & \multicolumn{2}{c}{$W$+1 jet} & \multicolumn{2}{c}{$W$+2 jets} & \multicolumn{2}{c}{$W$+3 jets} & \multicolumn{2}{c|}{$W$+$\ge$4 jets} & \multicolumn{2}{c}{$W$+1 jet} & \multicolumn{2}{c}{$W$+2 jets} & \multicolumn{2}{c}{$W$+3 jets} & \multicolumn{2}{c}{$W$+$\ge$4 jets}\\
\hline
\multicolumn{17}{c}{Single tag probabilities (\%)} \\ [2.4pt] \hline
$W$+light & 0.40 & 0.01 & 0.64 & 0.02 & 0.90 & 0.05 & 1.37 & 0.14 &
0.39 & 0.01 & 0.62 & 0.02 & 0.89 & 0.05 & 1.23 & 0.14 \\ \hline
$W$+light & 0.39 & 0.01 & 0.62 & 0.04 & 0.90 & 0.02 & 1.32 & 0.06 & 0.41 & 0.01 & 0.74 & 0.04 & 0.92 & 0.03 & 1.23 & 0.05 \\
$W(c\bar{c})$ & 9.3 & 0.1 & 8.6 & 0.3 & 8.9 & 0.2 & 9.2 & 0.9 & 9.4 & 0.1 & 9.2 & 0.2 & 8.6 & 0.1 & 10.2 & 0.7 \\
$W(b\bar{b})$ & 38.4 & 0.4 & 35.4 & 0.6 & 34.5 & 0.4 & 34.9 & 1.9 & 38.5 & 0.4 & 36.3 & 0.6 & 33.7 & 0.4 & 35.8 & 1.5 \\
$Wc$ & 9.6 & 0.1 & 9.6 & 0.2 & 9.7 & 0.3 & 10.2 & 0.3 & 9.6 & 0.1 & 9.4 & 0.2 & 9.4 & 0.3 & 9.7 & 0.3 \\
$Wc\bar{c}$ & \multicolumn{2}{c}{}& 15.6 & 0.4 & 14.8 & 1.1 & 16.4 & 0.6 & \multicolumn{2}{c}{}& 16.0 & 0.4 & 16.2 & 0.7 & 16.3 & 0.6 \\
$Wb\bar{b}$ & \multicolumn{2}{c}{}& 43.8 & 0.7 & 45.6 & 0.9 & 44.5 & 0.9 & \multicolumn{2}{c}{}& 44.0 & 0.8 & 44.0 & 1.0 & 44.0 & 0.8 \\
\hline
$W$+jets & 1.23 & 0.01 & 2.66 & 0.04 & 3.59 & 0.05 & 5.03 & 0.07 & 1.25 & 0.01 & 2.78 & 0.04 & 3.57 & 0.04 & 4.97 & 0.08 \\
\hline
\multicolumn{17}{c}{Double tag probabilities (\%)} \\ [2.4pt] \hline
$W$+light & \multicolumn{2}{c}{}& \multicolumn{2}{c}{$<0.01$} & \multicolumn{2}{c}{$<0.01$} & \multicolumn{2}{c|}{$<0.01$} & \multicolumn{2}{c}{}& \multicolumn{2}{c}{$<0.01$} & \multicolumn{2}{c}{$<0.01$} & \multicolumn{2}{c}{$<0.01$} \\
$W(c\bar{c})$ & \multicolumn{2}{c}{}& 0.03 & 0.01 & 0.09 & 0.01 & 0.14 & 0.05 & \multicolumn{2}{c}{}& 0.04 & 0.01 & 0.08 & 0.01 & 0.14 & 0.04 \\
$W(b\bar{b})$ & \multicolumn{2}{c}{}& 0.49 & 0.09 & 0.97 & 0.09 & 0.52 & 0.11 & \multicolumn{2}{c}{}& 0.96 & 0.15 & 0.77 & 0.07 & 1.35 & 0.39 \\
$Wc$ & \multicolumn{2}{c}{}& 0.023 & 0.002 & 0.052 & 0.004 & 0.082 & 0.004 & \multicolumn{2}{c}{}& 0.030 & 0.002 & 0.051 & 0.004 & 0.074 & 0.004 \\
$Wc\bar{c}$ & \multicolumn{2}{c}{}& 0.76 & 0.04 & 0.75 & 0.10 & 0.97 & 0.08 & \multicolumn{2}{c}{}& 0.80 & 0.04 & 0.94 & 0.10 & 1.05 & 0.09 \\
$Wb\bar{b}$ & \multicolumn{2}{c}{}& 12.2 & 0.5 & 13.1 & 0.8 & 14.1 & 0.6 & \multicolumn{2}{c}{}& 13.0 & 0.4 & 12.5 & 0.7 & 12.8 & 0.5 \\
\hline
$W$+jets & \multicolumn{2}{c}{}& 0.17 & 0.01 & 0.32 & 0.02 & 0.48 & 0.02 & \multicolumn{2}{c}{}& 0.19 & 0.01 & 0.31 & 0.01 & 0.47 & 0.02 \\
\hline \hline
\end{tabular}
\caption{Tagging probabilities (\%) for preselected \wplus
events for single tags (top rows) and double tags (bottom rows). The
uppermost row labeled $W$+light corresponds to the efficiencies
obtained from applying the light tagging efficiency parameterization to
the preselected signal sample. The rows labeled
$W$+jets summarize the average event tagging probabilities for
$W$ boson events. These values are not used in the analysis and are
included for informational purposes only. In all cases,
statistical uncertainties only are quoted.}
\label{tab:svt_tag_prob}
\end{table*}

\subsection{Evaluation of the QCD Multijet Background}
\label{sec:matrix_btag}
The QCD multijet background is evaluated by
applying the matrix method directly to the tagged samples.
Equation~\ref{eq:matrix}, originally defined for the
      preselected data in Sec.~\ref{sec:matrixmethod}, can be re-written
      for the single and double tagged samples and
      directly solved to obtain the number of QCD multijet events
      in the tagged samples.
	The rate at which a loose lepton in QCD multijet events
appears to be tight is remeasured for the tagged samples
and found to agree with the one used for the preselected samples.

As a cross check, the QCD multijet
background in the single tagged \eplus sample is obtained by multiplying
the number of QCD multijet events in the preselected sample ($N_{t}^{\text{QCD}}$)
by the corresponding average event tagging probability
$P^{\text{tag}}_{\text{QCD}}$, defined as the fraction of
tagged events in the loose-minus-tight \eplus sample.
The estimated number of tagged events is then given by
\begin{eqnarray*}
N^{\text{tag}}_{\text{QCD}} = N_{t}^{\text{QCD}} P^{\text{tag}}_{\text{QCD}} \,.
\end{eqnarray*}
Good agreement is observed between the matrix method and the cross check.

The cross check assumes that the heavy flavor composition in the
loose-minus-tight data sample, where the average event tagging probability is
derived, is identical to the heavy flavor composition of the QCD
multijet background in the preselected sample.
In the $e$+jets channel this assumption
applies, since the instrumental background mainly originates from
electromagnetically fluctuating jets misreconstructed as electrons.
In the \muplus channel however, the instrumental
background originates mainly from semileptonically decaying $b$
quarks to muons; the heavy flavor fraction is therefore enriched when the isolation
criteria is inverted, leading to a higher event tagging probability.
As the cross check cannot be applied to the \muplus channel,
results from the matrix method are used to extract the cross section
in both the \eplus and the \muplus channel.

Tables~\ref{tab:svt_summary_table_1}
and~\ref{tab:svt_summary_table_2} summarize the expected number of
QCD multijet events as a function of jet multiplicity
for single and double tag events, respectively.

\subsection{Physics Backgrounds}
\label{sec:smallbgr_btag}
Additional low rate electroweak processes that contribute to the
tagged sample are
diboson production ($WW \rightarrow l+\text{jets}$, $WZ \rightarrow l+\text{jets}$,
$WZ\rightarrow jjl\bar{l}$, $ZZ\rightarrow l\bar{l}jj$),
single top quark $s$- and $t$-channel production, and
$Z/\gamma^* \rightarrow \tau\tau \rightarrow l+\text{jets}$, where one $\tau$ decays
leptonically and the second one hadronically.
The $Z$+jets background where one of the two leptons is not reconstructed is
found to be negligible.

For a given process $i$, the number of events before tagging is determined as
\begin{eqnarray*}
{N}_{\text{bkg}\;i}^{\text{presel}} &=&
\sigma_{i} \varepsilon^{\text{presel}}_{i} Br_{i} {\mathcal L},
\end{eqnarray*}
where $\sigma_{i}$, $Br_{i}$, and $\mathcal{L}$ stand, respectively,
for the cross section,
branching ratio, and integrated luminosity for the process under consideration.
$\varepsilon^{\text{presel}}_{i}$ includes the
trigger efficiency for events that
pass the preselection and is obtained by folding into the MC
the per-lepton and per-jet trigger efficiencies measured in data.
The preselection efficiency is entirely determined from MC with the
appropriate scale factors applied. The estimated number of tagged
events is given by ${N}_{\text{bkg}\;i}^{\text{tag}} = {N}_{\text{bkg}\;i}^{\text{presel}} P^{\text{tag}}_{i}$,
with $P^{\text{tag}}_{i}$ the average event tagging probability for
the corresponding process.

Tables~\ref{tab:svt_summary_table_1}
and~\ref{tab:svt_summary_table_2} summarize the expected number of
events for each of the processes considered
 as a function of jet multiplicity
for single and double tag events, respectively.

\subsection{Observed and Predicted Numbers of Tagged Events}
\label{sec:obsPred}

\begin{figure*}
\begin{center}
\begin{tabular}{cc}
\mbox{\epsfig{file=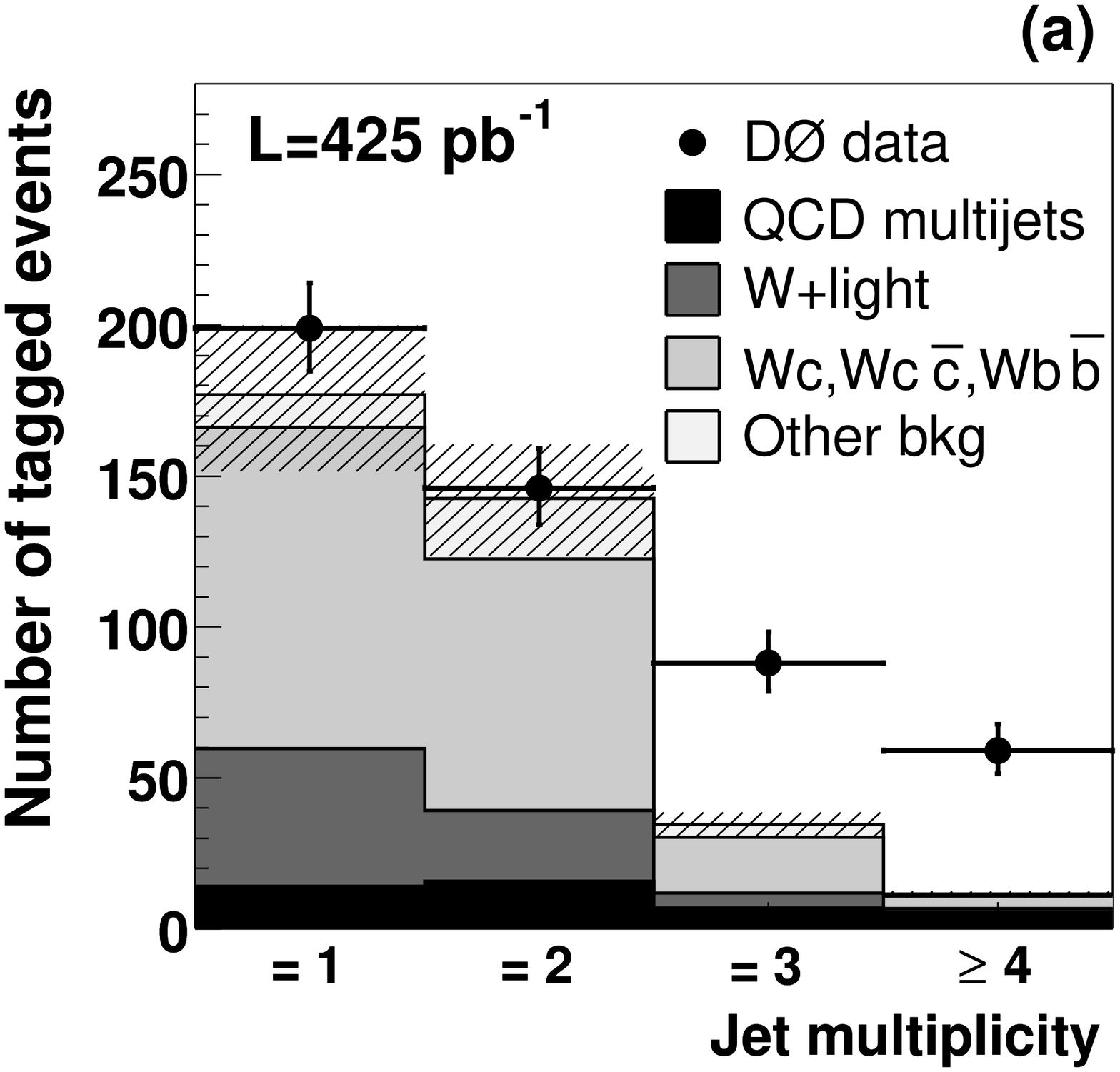,width=9cm}}
&
\mbox{\epsfig{file=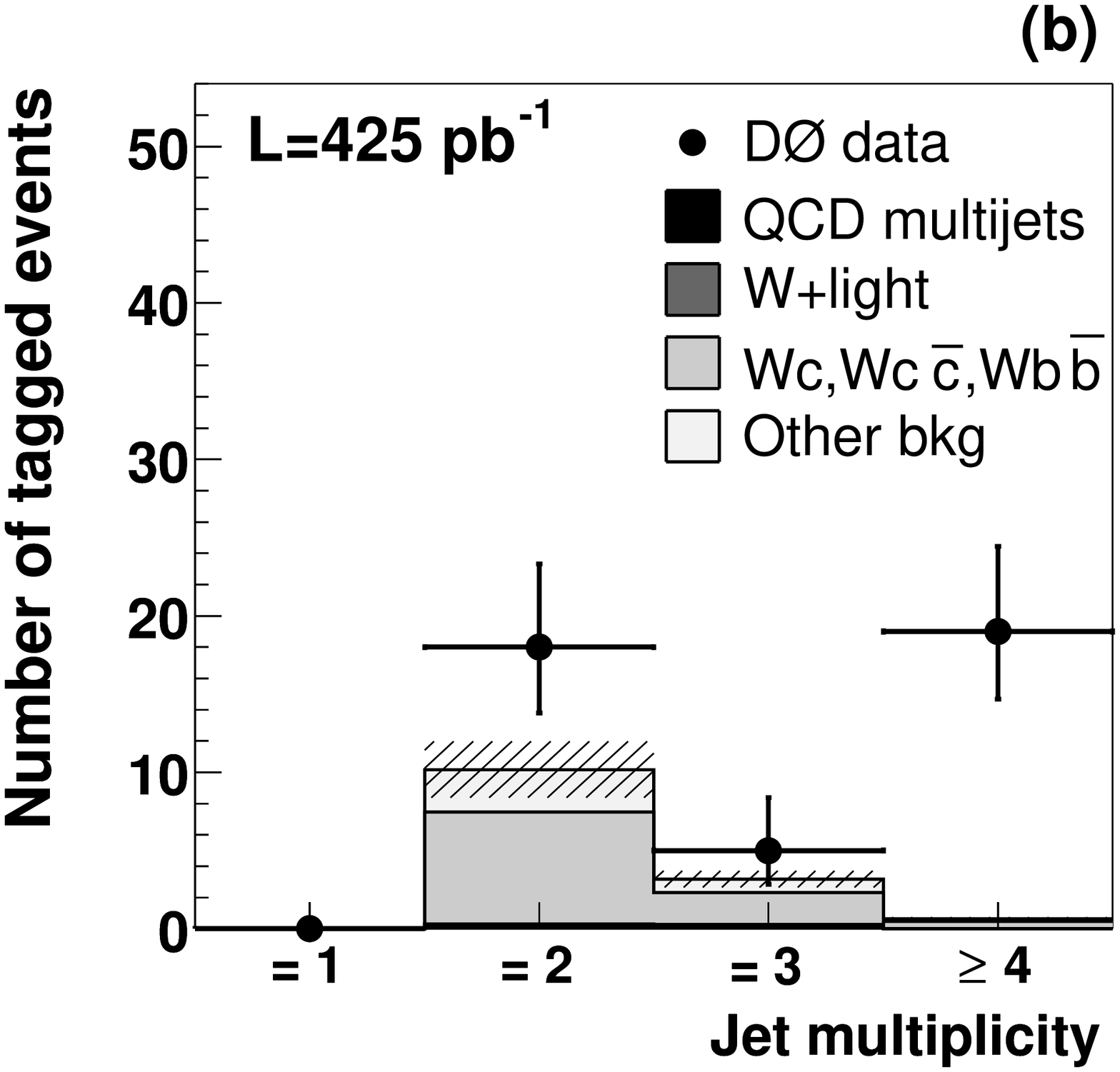,width=9cm}}
\end{tabular}
\end{center}
\caption{Observed number of tagged events in data compared
to the total SM background predictions (excluding \ttbar) for
(a) single tagged events and (b) double tagged events. The total
uncertainty on the background prediction is represented by the
hatched band. The excess of observed events in the third and fourth jet
multiplicity bins is attributed to \ttbar production.}
\label{fig:svt_summary_plot}
\end{figure*}

\begin{figure*}
\begin{center}
\begin{tabular}{cc}
\mbox{\epsfig{file=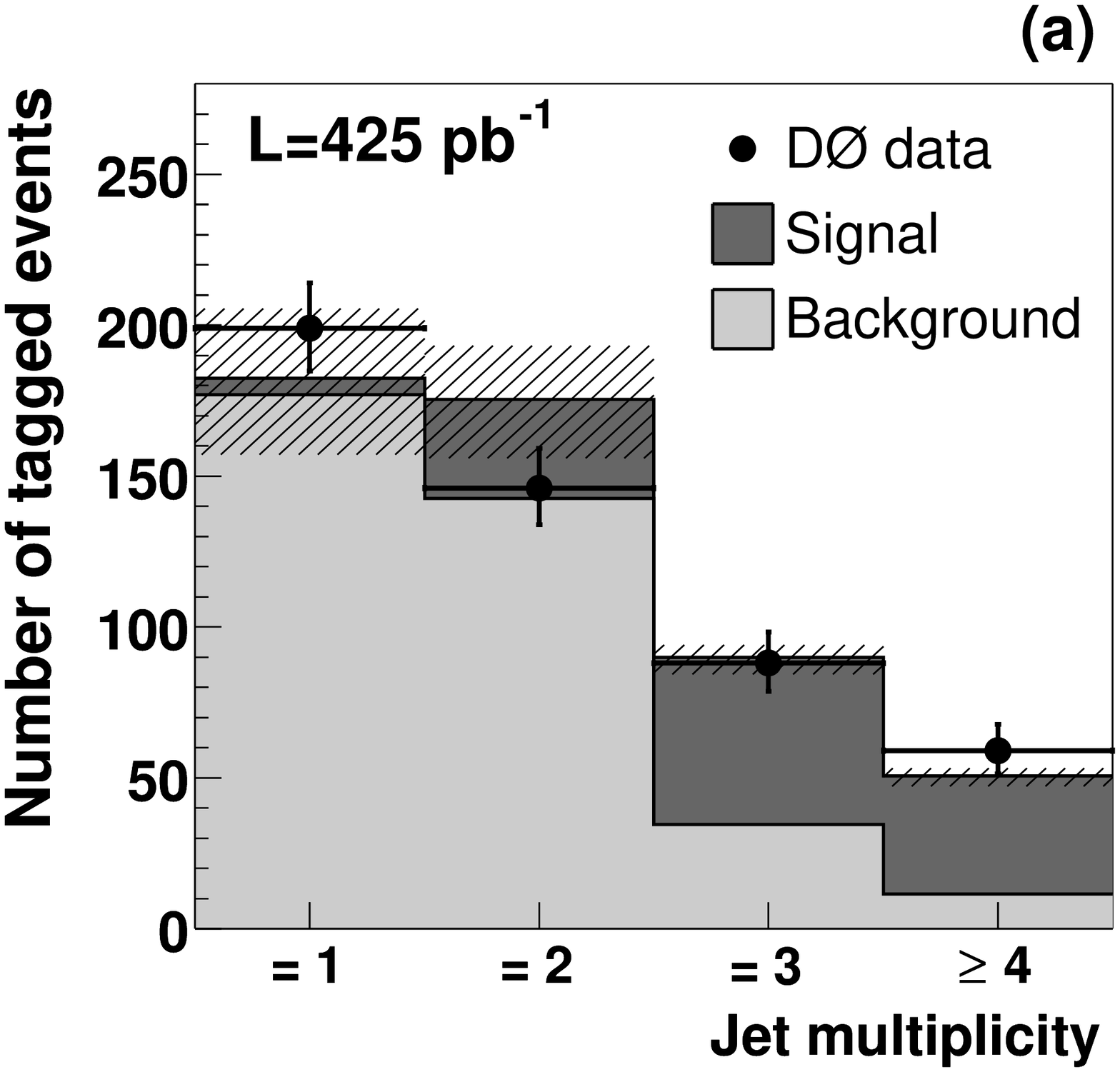,width=9cm}}
&
\mbox{\epsfig{file=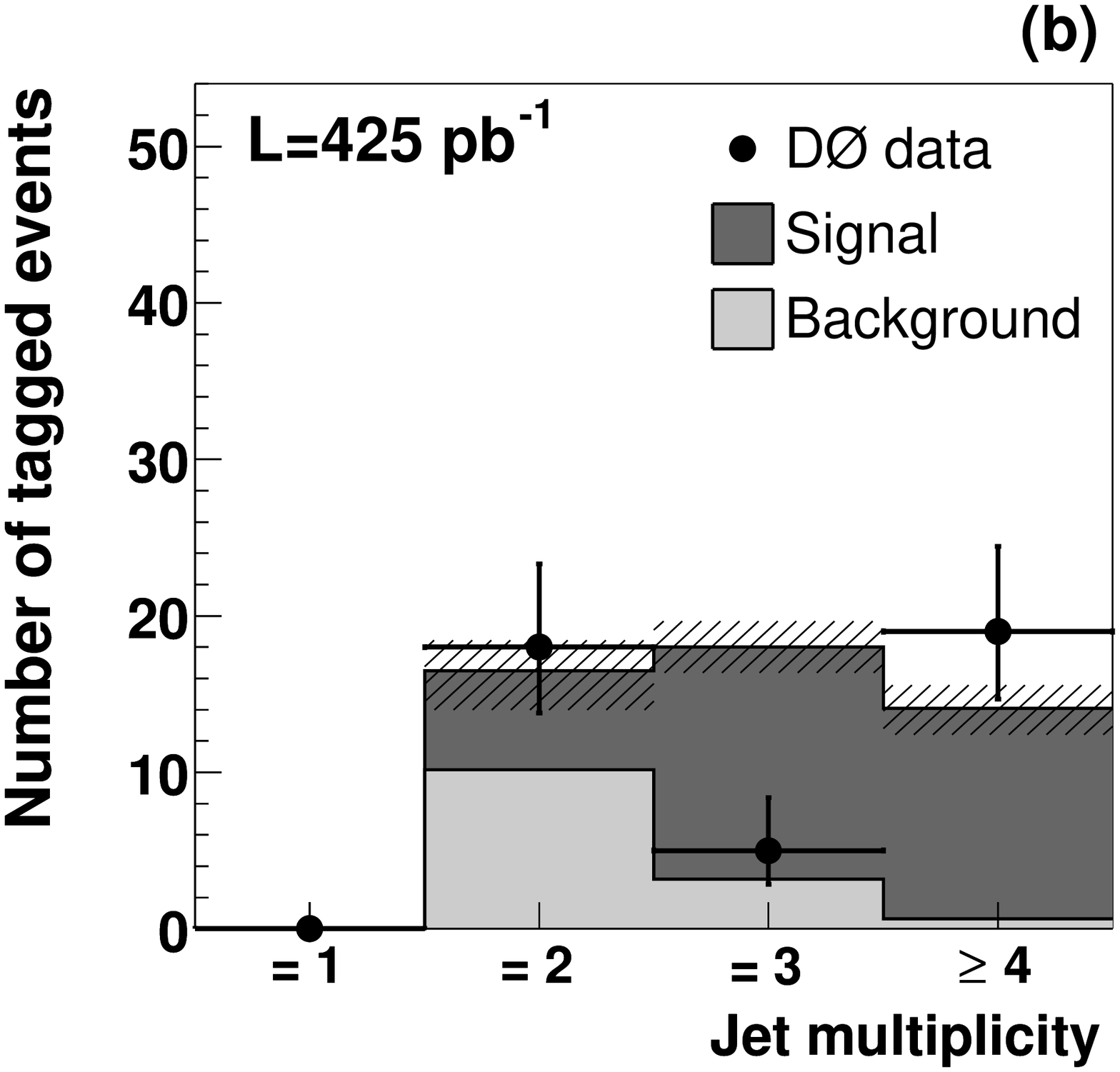,width=9cm}}
\end{tabular}
\end{center}
\caption{Observed number of tagged events in data compared
to the total SM prediction for
(a) single tagged events and (b) double tagged events. The number of 
\ttbar\ events shown is calculated assuming a cross section of 
$6.6\;\rm pb$. The total
uncertainty is represented by the
hatched band.}
\label{fig:svt_summary_plot2}
\end{figure*}

The numbers of observed and predicted single and double tagged events
are summarized in Tables~\ref{tab:svt_summary_table_1}
and~\ref{tab:svt_summary_table_2}, respectively.
Figure~\ref{fig:svt_summary_plot}
shows the observed number of tagged events in data compared
to the total SM background predictions, excluding \ttbar.
The background in the first jet multiplicity bin is dominated by $W$+light
and $Wc$ events. The contribution from heavy flavor production,
particularly from $Wb\bar{b}$, dominates for events with three or more jets.
Very good agreement between observation and background prediction is
observed in the background-dominated first and second jet multiplicity bins,
which gives confidence in the background estimate of the analysis.
A clear excess of observed events over background is seen in the third
and fourth jet multiplicity bins. The excess events are attributed to \ttbar
production and are used to extract the cross section.
Figure~\ref{fig:svt_summary_plot2}
shows the observed number of tagged events in data compared
to the total SM predictions including \ttbar.
The number of
\ttbar events shown is calculated based on the measured
cross section.

\begin{table*}[htbp]
\centering
\begin{tabular}{l|r@{$\pm$}lr@{$\pm$}lr@{$\pm$}lr@{$\pm$}l|r@{$\pm$}lr@{$\pm$}lr@{$\pm$}lr@{$\pm$}l}
\hline  \hline
&\multicolumn{8}{c|}{$e$+jets}&\multicolumn{8}{c}{$\mu$+jets}\\
\hline
 & \multicolumn{2}{c}{1 jet} & \multicolumn{2}{c}{2 jets} & \multicolumn{2}{c}{3 jets} & \multicolumn{2}{c|}{$\ge$4 jets} & \multicolumn{2}{c}{1 jet} & \multicolumn{2}{c}{2 jets} & \multicolumn{2}{c}{3 jets} & \multicolumn{2}{c}{$\ge$4 jets}\\
\hline
$W$+light & 20.9 & 0.7 & 10.1 & 0.7 & 2.45 & 0.19 & 0.59 & 0.13 & 24.9 & 0.8 & 13.2 & 0.8 & 2.63 & 0.19 & 0.46 & 0.11 \\
$W(c\bar{c})$ & 6.6 & 0.1 & 3.7 & 0.2 & 0.88 & 0.06 & 0.26 & 0.06 & 7.5 & 0.1 & 4.3 & 0.1 & 0.90 & 0.06 & 0.24 & 0.06 \\
$W(b\bar{b})$ & 18.8 & 0.3 & 9.6 & 0.3 & 2.25 & 0.16 & 0.58 & 0.12 & 21.6 & 0.4 & 10.9 & 0.3 & 2.32 & 0.15 & 0.50 & 0.12 \\
$Wc$ & 24.3 & 0.5 & 11.2 & 0.4 & 1.53 & 0.12 & 0.24 & 0.05 & 27.6 & 0.5 & 12.0 & 0.4 & 1.56 & 0.11 & 0.19 & 0.04 \\
$Wc\bar{c}$ & \multicolumn{2}{c}{}& 4.9 & 0.2 & 1.39 & 0.15 & 0.40 & 0.09 & \multicolumn{2}{c}{}& 5.6 & 0.2 & 1.62 & 0.13 & 0.34 & 0.08 \\
$Wb\bar{b}$ & \multicolumn{2}{c}{}& 10.1 & 0.3 & 3.00 & 0.22 & 0.70 & 0.15 & \multicolumn{2}{c}{}& 11.1 & 0.3 & 3.05 & 0.21 & 0.58 & 0.13 \\
\hline
$W$+jets & 70.6 & 0.9 & 49.6 & 0.9 & 11.5 & 0.4 & 2.77 & 0.26 & 81.6 & 1.0 & 57.1 & 1.0 & 12.1 & 0.4 & 2.31 & 0.23 \\
\hline
QCD & 6.8 & 1.5 & 10.0 & 1.7 & 5.2 & 1.2 & 2.95 & 0.98 & 7.2 & 1.3 & 5.8 & 1.3 & 1.57 & 0.89 & 2.77 & 1.02 \\
\hline
Single top & 3.30 & 0.07 & 7.3 & 0.1 & 1.88 & 0.06 & 0.30 & 0.03 & 2.65 & 0.05 & 6.5 & 0.1 & 1.72 & 0.04 & 0.27 & 0.02 \\
Diboson & 2.26 & 0.10 & 2.75 & 0.11 & 0.23 & 0.03 & \multicolumn{2}{c|}{$<0.01$} & 2.28 & 0.10 & 2.94 & 0.11 & 0.22 & 0.03 & \multicolumn{2}{c}{$<0.01$} \\
$Z/\gamma^*\rightarrow\tau^+\tau^-$ & 0.15 & 0.04 & 0.40 & 0.07 & 0.03 & 0.01 & \multicolumn{2}{c|}{$<0.01$} & 0.19 & 0.07 & 0.29 & 0.05 & 0.09 & 0.05 & 0.01 & 0.02 \\
\hline
$N_{\text{bkg}}$ & 83.1 & 1.7 & 70.1 & 2.0 & 18.8 & 1.4 & 6.0 & 1.1 & 93.9 & 1.7 & 72.6 & 1.7 & 15.7 & 1.1 & 5.4 & 1.1 \\
Syst. & \multicolumn{2}{c}{+10.7$-$11.8} & \multicolumn{2}{c}{+8.5$-$9.0} & \multicolumn{2}{c}{+1.9$-$2.0} & \multicolumn{2}{c|}{+0.5$-$0.5} & \multicolumn{2}{c}{+12.2$-$13.4} & \multicolumn{2}{c}{+9.3$-$9.9} & \multicolumn{2}{c}{+2.0$-$2.1} & \multicolumn{2}{c}{+0.4$-$0.4} \\
\hline
$t\bar{t}\rightarrow l$+jets & 1.07 & 0.18 & 11.7 & 0.3 & 27.3 & 0.4 & 19.8 & 0.3 & 0.60 & 0.19 & 8.0 & 0.4 & 23.6 & 0.4 & 18.8 & 0.4 \\
$t\bar{t}\rightarrow ll$ & 2.28 & 0.04 & 7.1 & 0.1 & 2.34 & 0.04 & 0.33 & 0.02 & 1.60 & 0.03 & 5.9 & 0.1 & 2.18 & 0.04 & 0.29 & 0.01 \\
\hline
$N_{\text{pred}}$ & 86.5 & 1.7 & 88.9 & 2.0 & 48.5 & 1.4 & 26.2 & 1.1 & 96.1 & 1.7 & 86.5 & 1.7 & 41.5 & 1.1 & 24.5 & 1.1 \\
Syst. & \multicolumn{2}{c}{+10.7$-$11.9} & \multicolumn{2}{c}{+8.3$-$10.4} & \multicolumn{2}{c}{+2.0$-$3.3} & \multicolumn{2}{c|}{+1.0$-$3.5} & \multicolumn{2}{c}{+12.3$-$13.4} & \multicolumn{2}{c}{+9.8$-$9.8} &
\multicolumn{2}{c}{+2.2$-$2.5} & \multicolumn{2}{c}{+2.6$-$1.0} \\
\hline
$N_{\text{obs}}$ & \multicolumn{2}{c}{94} & \multicolumn{2}{c}{78} & \multicolumn{2}{c}{47} & \multicolumn{2}{c|}{33} & \multicolumn{2}{c}{105} & \multicolumn{2}{c}{68} & \multicolumn{2}{c}{41} & \multicolumn{2}{c}{26} \\
\hline \hline
\end{tabular}
\caption{Summary of observed ($N_{\text{obs}}$) and
predicted ($N_{\text{pred}})$ numbers of single tagged events in
the \eplus and the \muplus channels. Uncertainties shown are statistical;
the systematic uncertainties are included in the row labeled Syst. 
The number of \ttbar events quoted is calculated assuming a cross section
of $6.6\;\rm pb$.}
\label{tab:svt_summary_table_1}
\end{table*}

\begin{table*}[htbp]
\centering
\begin{tabular}{l|r@{$\pm$}lr@{$\pm$}lr@{$\pm$}l|r@{$\pm$}lr@{$\pm$}lr@{$\pm$}l}
\hline \hline
&\multicolumn{6}{c|}{$e$+jets}&\multicolumn{6}{c}{$\mu$+jets}\\
\hline
 & \multicolumn{2}{c}{2 jets} & \multicolumn{2}{c}{3 jets} & \multicolumn{2}{c|}{$\ge$4 jets} & \multicolumn{2}{c}{2 jets} & \multicolumn{2}{c}{3 jets} & \multicolumn{2}{c}{$\ge$4 jets}\\
\hline
$W$+light & 0.017 & 0.003 & \multicolumn{2}{c}{$<0.01$} & \multicolumn{2}{c|}{$<0.01$} & 0.027 & 0.003 & \multicolumn{2}{c}{$<0.01$} & \multicolumn{2}{c}{$<0.01$} \\
$W(c\bar{c})$ & 0.014 & 0.002 & \multicolumn{2}{c}{$<0.01$} & \multicolumn{2}{c|}{$<0.01$} & 0.019 & 0.003 & \multicolumn{2}{c}{$<0.01$} & \multicolumn{2}{c}{$<0.01$} \\
$W(b\bar{b})$ & 0.13 & 0.03 & 0.06 & 0.01 & \multicolumn{2}{c|}{$<0.01$} & 0.29 & 0.05 & 0.05 & 0.01 & 0.02 & 0.01 \\
$Wc$ & 0.027 & 0.002 & \multicolumn{2}{c}{$<0.01$} & \multicolumn{2}{c|}{$<0.01$} & 0.039 & 0.003 & \multicolumn{2}{c}{$<0.01$} & \multicolumn{2}{c}{$<0.01$} \\
$Wc\bar{c}$ & 0.24 & 0.01 & 0.07 & 0.01 & 0.02 & 0.01 & 0.28 & 0.01 & 0.09 & 0.01 & 0.02 & 0.01 \\
$Wb\bar{b}$ & 2.80 & 0.13 & 0.86 & 0.08 & 0.22 & 0.05 & 3.30 & 0.14 & 0.87 & 0.07 & 0.17 & 0.04 \\
\hline
$W$+jets & 3.23 & 0.13 & 1.00 & 0.08 & 0.26 & 0.05 & 3.96 & 0.15 & 1.02 & 0.08 & 0.22 & 0.04 \\
\hline
QCD & \multicolumn{2}{c}{$<0.01$} & 0.27 & 0.22 & \multicolumn{2}{c|}{$<0.01$} & 0.26 & 0.29 & \multicolumn{2}{c}{$<0.01$} & \multicolumn{2}{c}{$<0.01$} \\
\hline
Single top & 1.07 & 0.02 & 0.39 & 0.02 & 0.07 & 0.01 & 0.93 & 0.01 & 0.37 & 0.01 & 0.07 & 0.01 \\
Diboson & 0.34 & 0.02 & 0.04 & 0.01 & \multicolumn{2}{c|}{$<0.01$} & 0.26 & 0.02 & 0.03 & 0.01 & \multicolumn{2}{c}{$<0.01$} \\
$Z\rightarrow\tau^+\tau^-$ & \multicolumn{2}{c}{$<0.01$} & \multicolumn{2}{c}{$<0.01$} & \multicolumn{2}{c|}{$<0.01$} & \multicolumn{2}{c}{$<0.01$} & 0.02 & 0.02 & \multicolumn{2}{c}{$<0.01$} \\
\hline
$N_{\text{bkg}}$ & 4.64 & 0.28 & 1.70 & 0.40 & 0.34 & 0.29 & 5.42 & 0.33 & 1.44 & 0.34 & 0.29 & 0.38 \\
Syst. & \multicolumn{2}{c}{+0.83$-$0.81} & \multicolumn{2}{c}{+0.26$-$0.25} & \multicolumn{2}{c|}{+0.06$-$0.06} & \multicolumn{2}{c}{+0.99$-$0.97} & \multicolumn{2}{c}{+0.27$-$0.25} & \multicolumn{2}{c}{+0.05$-$0.06} \\
\hline
$t\bar{t}\rightarrow l$+jets & 1.72 & 0.19 & 7.3 & 0.3 & 6.9 & 0.2 & 1.02 & 0.15 & 6.2 & 0.3 & 6.3 & 0.3 \\
$t\bar{t}\rightarrow ll$ & 1.81 & 0.02 & 0.65 & 0.01 & 0.09 & 0.01 & 1.50 & 0.02 & 0.61 & 0.01 & 0.08 & 0.01 \\
\hline
$N_{\text{pred}}$ & 8.2 & 0.3 & 9.7 & 0.4 & 7.3 & 0.3 & 7.9 & 0.4 & 8.3 & 0.3 & 6.7 & 0.4 \\
Syst. & \multicolumn{2}{c}{+0.8$-$1.9} & \multicolumn{2}{c}{+0.6$-$1.3} & \multicolumn{2}{c|}{+0.4$-$1.8} & \multicolumn{2}{c}{+1.3$-$1.0} & \multicolumn{2}{c}{+1.3$-$0.7} & \multicolumn{2}{c}{+1.7$-$0.4} \\
\hline
$N_{\text{obs}}$ & \multicolumn{2}{c}{12} & \multicolumn{2}{c}{2} & \multicolumn{2}{c|}{11} & \multicolumn{2}{c}{6} & \multicolumn{2}{c}{3} & \multicolumn{2}{c}{8}\\
\hline \hline
\end{tabular}
\caption{Summary of observed ($N_{\text{obs}}$) and
predicted ($N_{\text{pred}})$ numbers of double tagged events in
the \eplus and the \muplus channels. Uncertainties shown are statistical;
the systematic uncertainties are included in the row labeled Syst.
The number of \ttbar events quoted is calculated assuming a cross section
of $6.6\;\rm pb$.}
\label{tab:svt_summary_table_2}
\end{table*}

\section{Cross Section Result}
\label{sec:xs}
The \ttbar production cross section is extracted from the excess of
tagged events over background expectation according to:
\begin{eqnarray*}
\sigma = \frac {N_{\text{obs}} - N_{\text{bkg}}}
{Br \cdot {\mathcal L}
\cdot \varepsilon_{\text{presel}} \cdot P^{\text{tag}}} \;,
\end{eqnarray*}
where $Br$ is the branching ratio of the considered final state, $\mathcal L$
is the integrated luminosity, $\varepsilon_{\text{presel}}$ is the
\ttbar preselection efficiency, and $P^{\text{tag}}$ is the probability for a
\ttbar event to have one or more jets identified as $b$ jets.

The \ttbar production cross section is calculated by performing a
maximum likelihood fit to the observed number of events. The analysis
is split into eight different channels:
$e$+3 jets single tag, $e$+3 jets double tag,
$e$+4 jets single tag, $e$+4 jets double tag,
$\mu$+3 jets single tag, $\mu$+3 jets double tag,
$\mu$+4 jets single tag, and $\mu$+4 jets double tag.
The resulting cross sections are given  for the
electron and the muon channels separately and combined. If the index $\gamma$
refers to one of the eight channels,
the likelihood
${\mathcal L}_{1}$ to observe $N^{\text{obs}}_{\gamma}$
for a cross section $\sigma_{t\bar t}$
is proportional to
\begin{eqnarray}
  \label{eq:poisson1} {\mathcal L}_{1}  & = & \prod_{\gamma} {\mathcal
  P}[N^{\mathrm{ obs}}_{\gamma},N^{\mathrm{pred}}_{\gamma}(\sigma_{t\bar t})]
  \,.
\end{eqnarray}
${\mathcal P}(n,\mu)=\frac{\mu^n\mathrm{e}^{-\mu}}{n\mathrm{!}}$
generically
denotes the Poisson probability function for $n$ observed
events, given an expectation of $\mu$~events.
The predicted number of events in each channel is the sum of the predicted
number of background events and the number of expected
\ttbar events. Both the number of \wplus events before tagging and
the number of expected \ttbar events are functions of
the \ttbar cross section that is being determined.
For each iteration of the maximization procedure of the likelihood,
the number of \ttbar events in the untagged sample is calculated
and the number of \wplus is  rederived.
A detailed explanation of the treatment of the event statistics
in the cross section calculation can be found in Appendix~\ref{sec:stat_xs}.

The final cross section is determined using a nuisance parameter
likelihood method~\cite{nuisance} that incorporates all systematic uncertainties in the fit
in such a way that allows them to affect the central value of the cross
section. In this approach, each independent source of systematic uncertainty
is modeled by a free
parameter. Each nuisance parameter
is modeled with a Gaussian centered on zero and with a standard
deviation of one. The nuisance parameters are allowed to change the central
values of all efficiencies, tagging probabilities, and flavor fractions, which
are allowed to vary within their uncertainties. The correlations
are taken into account in a natural way, by letting the same nuisance
parameter affect different variables.
The total likelihood function
that is maximized is the product of ${\mathcal L}_{1}$ and ${\mathcal L}_{2}$,
with
\begin{eqnarray*}
{\mathcal L}_{2}  = \prod_{i} {\mathcal G}(\nu_i;0,1) \,,
\end{eqnarray*}
where ${\mathcal G}(\nu_i;0,1)$ is the normal probability of the nuisance parameter $i$
to take the value $\nu_i$.

The measured $t\bar{t}$
production cross sections for a top quark mass of $175\;\rm GeV$ are
\begin{eqnarray*}
\mu+{\rm jets}&:&\quad \sigma_{t\overline{t}} = \nonumber
   6.1^{+1.3}_{-1.2}{\rm (stat+syst)}\:\pm 0.4  \:{\rm(lum)}\:{\rm pb},\\ \nonumber
e+{\rm jets}&:&\quad \sigma_{t\overline{t}} =
      6.9^{+1.4}_{-1.2}{\rm (stat+syst)}\:\pm 0.4 \:{\rm(lum)} \:{\rm pb},\\ \nonumber
l+{\rm jets} &:&\quad \sigma_{t\overline{t}} =
   6.6\pm 0.9{\rm (stat+syst)}\:\pm 0.4 \:{\rm(lum)}\:{\rm pb}. \nonumber
\end{eqnarray*}
The first uncertainty corresponds to the combined statistical and
systematic uncertainties, and the second one to the luminosity error of $\pm6.1\%$.

A complete list of systematic uncertainties is given
in Table~\ref{tab:table_systematics},
where a cross indicates if the
background normalization ($\Delta b$) and/or the $t\bar{t}$ efficiency
($\Delta \varepsilon$) are affected within a given channel.
The systematic uncertainties have been classified as
uncorrelated (usually of statistical origin in either
MC or data) or correlated. The correlation can
be between channels (i.e. \eplus and \muplus)
and/or between jet multiplicity bins
($N_{\text{jet}}=3$ and $N_{\text{jet}}\geq 4$) within a particular channel.
All systematic uncertainties are fully correlated between
the single and double tagged samples.

\begin{table}[!htpb]
  \begin{center}
  \begin{tabular}{cl|c|c|c|c}
  \hline \hline
    \multicolumn{2}{c|}{} & \multicolumn{2}{c|}{e+jets} & \multicolumn{2}{c}{$\mu$+jets} \\
        \cline{3-6}
    \multicolumn{2}{c|}{} & $\Delta b$ & $\Delta \varepsilon$ & $\Delta b$ & $\Delta \varepsilon$ \\
        \hline
        \begin{picture}(0,0)(10,60)\rotatebox{90}{Uncorrelated} \end{picture}
    & { Muon trigger}&          &            & $\times$ &  $\times$\\
    & { EM trigger}& $\times$     & $\times$  &          &  \\
    & { Muon preselection} &          &            & $\times$ &  $\times$\\
    & { Electron preselection}& $\times$ & $\times$  &          & \\
    & { Preselection efficiency (MC statistics)} &          & $\times$ &          & $\times$ \\
    & { $\varepsilon_{\text{QCD}}$ and $\varepsilon_{\text{sig}}$} & $\times$ &    & $\times$ &  \\
    & { Matrix method (data statistics)} & $\times$ & $\times$   & $\times$ & $\times$ \\
    & { $W$ fractions (MC statistics)}             & $\times$ &          & $\times$ &          \\
    \hline
        \begin{picture}(0,0)(10,50)\rotatebox{90}{Correlated} \end{picture}
    & { Jet trigger}                      & $\times$ & $\times$ & $\times$ &
    $\times$\\
    & { Jet preselection}                     & $\times$ & $\times$ & $\times$ & $\times$ \\
    & { Taggability in data}                 & $\times$ & $\times$ & $\times$ & $\times$ \\
    & { Flavor dependence of taggability}    & $\times$ & $\times$ & $\times$ & $\times$ \\
    & { Semileptonic $b$ tagging efficiency in data}        & $\times$ & $\times$ & $\times$ & $\times$ \\
    & { Semileptonic $b$ tagging efficiency in MC}      & $\times$ & $\times$ & $\times$ & $\times$ \\
    & { Inclusive $b$ tagging efficiency in MC}       & $\times$ & $\times$ & $\times$ & $\times$ \\
    & { Inclusive $c$ tagging efficiency in MC}       & $\times$ & $\times$ & $\times$ & $\times$ \\
    & { Negative tagging efficiency in data}        & $\times$ & $\times$ & $\times$ & $\times$ \\
    & { $SF_{ll}$ and $SF_{hf}$}               & $\times$ & $\times$ & $\times$ & $\times$ \\
    & { $W$ fractions }                  & $\times$ &          & $\times$ &          \\
    \hline \hline
  \end{tabular}
  \caption{Summary of systematic uncertainties affecting the signal
  efficiency and/or background prediction. The labels
  \emph{correlated} and \emph{uncorrelated}
  refer to the \muplus and \eplus channels.}
  \label{tab:table_systematics}
  \end{center}
\end{table}

The nuisance parameter likelihood provides the total uncertainty on the
cross section including contributions from systematic and statistical
origin.
To estimate the contribution of each individual systematic source,
all but the corresponding nuisance parameter are fixed
in the fit, and the
maximization is redone. The statistical contribution
is then deconvoluted from the
obtained uncertainty to extract the contribution for that particular source.
The resulting systematic uncertainties are summarized in
Table~\ref{tab:xs_svt_emu}.

The total uncertainty, excluding luminosity, is $\approx 14\% $.
The main contribution of $\approx 11\% $ is statistical; the
remaining $\approx 8\% $ is due to systematic effects. The primary contribution
to the systematic uncertainties arises from the semileptonic
$b$ tagging efficiency measured in data. The second largest source of
systematic uncertainty
originates from the matching of $W$ fractions and higher-order effects.

\begin{table}[htpb!]
  \centering
  \begin{tabular}{l|cc}
    \hline \hline
    Source    &   $\sigma^{+}$  & $\sigma^{-}$ \\
\hline
                                Muon trigger &    0.05 &   0.07 \\
                                  EM trigger &    0.00 &   0.01 \\
                                 Jet trigger &    0.00 &   0.01 \\   \hline
                           Muon preselection &    0.16 &   0.14 \\
                       Electron preselection &    0.17 &   0.15 \\
                            Jet preselection &    0.13 &   0.11 \\
     Preselection efficiency (MC statistics) &    0.06 &   0.04 \\   \hline
$\varepsilon_{\mathrm{QCD}}$ and $\varepsilon_{\mathrm{sig}}$ in \muplus channel & 0.04 & 0.03 \\
$\varepsilon_{\mathrm{QCD}}$ and $\varepsilon_{\mathrm{sig}}$ in \eplus channel  & 0.06 & 0.00 \\
             Matrix Method (data statistics) &    0.15 &   0.15 \\   \hline
                         Taggability in data &    0.03 &   0.00 \\
            Flavor dependence of taggability &    0.00 &   0.03 \\
 Semileptonic $b$ tagging efficiency in data &    0.33 &   0.24 \\
   Semileptonic $b$ tagging efficiency in MC &    0.17 &   0.04 \\
      Inclusive $b$ tagging efficiency in MC &    0.00 &   0.00 \\
      Inclusive $c$ tagging efficiency in MC &    0.01 &   0.00 \\
         Negative tagging efficiency in data &    0.00 &   0.01 \\
                     $SF_{ll}$ and $SF_{hf}$ &    0.01 &   0.00 \\   \hline
                               $W$ fractions &    0.29 &   0.27 \\
               $W$ fractions (MC statistics) &    0.03 &   0.03 \\   \hline
   Total systematics (quad sum of the above) &    0.57 &   0.47 \\   \hline
Total uncertainty (nuisance parameter lhood) &    0.94 &   0.86 \\
\hline \hline
  \end{tabular}
   \caption{Systematic uncertainties in the combined \lplus channel.
  }
  \label{tab:xs_svt_emu}
\end{table}

\begin{figure}[htbp]
\begin{center}
\mbox{\epsfig{file=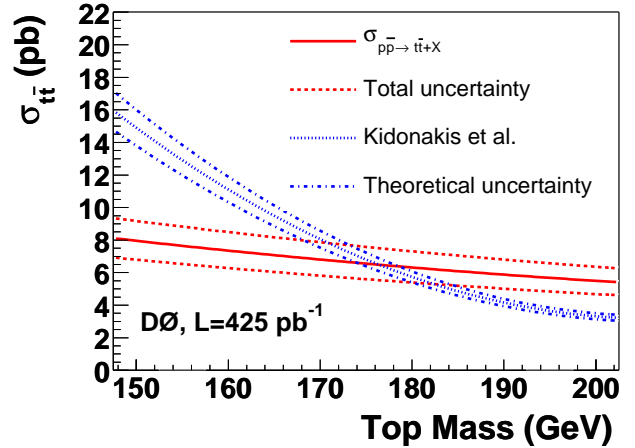,width=0.5\textwidth}}
\end{center}
\caption{Top quark mass dependence of the measured cross section
compared to the theoretical prediction \protect \cite{theoxsec}.}
\label{fig:XS_vs_mtop-svt}
\end{figure}

The measured cross section depends on the assumed mass of the top quark
$m_t$. The dependence was studied by repeating the analysis on MC \ttbar
samples generated at different values of $m_t$. The resulting
dependence can be
parameterized as
$\sigma_{t\overline{t}}(m_t) = 0.000273  m_t^2 - 0.145  m_t + 23.5 $ for
the central value, $-0.00704 m_t  + 2.26 $ for the 
$+1\sigma$ 
uncertainty, and $0.00687 m_t - 2.17$ for the 
$-1\sigma$ uncertainty. The dependence 
is shown in Fig.~\ref{fig:XS_vs_mtop-svt}.

%

\section{Conclusions}
\label{sec:conclu}
A measurement of the \ttbar production cross section in \ppbar
collisions at a \CME of 1.96~TeV is presented in events with a
lepton, a neutrino, and $\ge 3$ jets. After a preselection of
the objects in the final state, a lifetime
$b$ tagging algorithm which explicitly reconstructs secondary
vertices is applied, removing approximately 95\% of the background
while keeping 60\% of the \ttbar signal. The measurement combines
the \muplus and the \eplus channels, using 422~pb$^{-1}$ and
425~pb$^{-1}$ of data, respectively.
The measured \ttbar production cross section for a top quark mass
of $175\;\rm GeV$ is
\formule
{
\sigma_{p\overline{p}\rightarrow t\overline{t}+X} =
6.6\pm0.9\:{\rm (stat+syst)}\:
\pm0.4\:{\rm(lum)}\:{\rm pb}\,,\nonumber
}
in good agreement with SM expectations. The systematic uncertainty on the
result (excluding luminosity) is $\approx 8$\%. This represents a factor
of three reduction in the systematic uncertainty compared to previous
publications by the D0 collaboration~\cite{D0Run2}, making this
result the most precise D0 measurement of the \ttbar
production cross section to date.

%
We thank the staffs at Fermilab and collaborating institutions, 
and acknowledge support from the 
DOE and NSF (USA);
CEA and CNRS/IN2P3 (France);
FASI, Rosatom and RFBR (Russia);
CAPES, CNPq, FAPERJ, FAPESP and FUNDUNESP (Brazil);
DAE and DST (India);
Colciencias (Colombia);
CONACyT (Mexico);
KRF and KOSEF (Korea);
CONICET and UBACyT (Argentina);
FOM (The Netherlands);
PPARC (United Kingdom);
MSMT (Czech Republic);
CRC Program, CFI, NSERC and WestGrid Project (Canada);
BMBF and DFG (Germany);
SFI (Ireland);
The Swedish Research Council (Sweden);
Research Corporation;
Alexander von Humboldt Foundation;
and the Marie Curie Program.
%

\appendix

\section{Monte Carlo generation of \wplus events}
\label{sec:wsamples}

The \wplus background is simulated using {\alpgen}~{\scshape
1.3}~\cite{alpgen} followed by {\pythia}~{\scshape 6.2}~\cite{pythia}
to simulate the underlying event and the hadronization.
The samples are generated separately for processes with 1, 2, 3, and 4 or more
partons in the final state, as summarized in Table~\ref{tab:alpgen_xsect}.
No parton-level cuts are applied on the heavy quarks ($c$ or $b$) except for
the $c$ quark in the single $c$ quark production process; the correct masses for the
$c$ and the $b$ quark are also included.
The processes
$Wc\bar cc\bar c$, $Wb\bar bc\bar c$, and $Wb\bar bb\bar b$ are not
included as their cross sections are negligible.
$W$ bosons are forced to decay to leptons;
taus are subsequently forced to decay leptonically using
{\tauola}. The respective fraction of
$W{\rar}\tau\nu$ events is adjusted in the overall sample to correctly reflect
its contributions to the \eplus and \muplus channels.

\begin{table}[htpb]
\begin{tabular}{l r|l r|l r|l r} \hline \hline
Process &   $  \sigma$(pb) &
Process &   $  \sigma$(pb) &
Process &   $  \sigma$(pb) &
Process &   $  \sigma$(pb) \\\hline
$Wj$ & 1600  & $Wjj$ & 517 & $Wjjj$ & 163 & $Wjjjj$ & 49.5 \\
$Wc$ &  51.8 & $Wcj$ &   28.6 & $Wcjj$ &  19.4 & $Wcjjj$ & 3.15 \\
&& $Wb\bar{b}$ &  9.85 & $Wb\bar{b}J$ & 5.24 & $Wb\bar{b}Jj$ & 2.86 \\
&& $Wc\bar{c}$ & 24.3 & $Wc\bar{c}J$ & 12.5 & $Wc\bar{c}Jj$ & 5.83 \\\hline \hline
\end{tabular}
\caption{\wplus boson processes in {\alpgen} and their cross sections for
the leptonic $W$ boson decay, $\sigma \equiv
\sigma_{p\overline{p}\rightarrow W+\text{jets}}Br(W\rightarrow l \nu)$, where
\jet=\upq,\downq,\strangeq,\gluon and \Jet=\upq,\downq,\strangeq,\gluon,\charmq.}
\label{tab:alpgen_xsect}
\end{table}

The leading-order parton level calculations performed
by {\alpgen} need to be consistently combined with the partonic evolution
given by the shower MC
program {\pythia} to avoid the double counting of configurations
leading to the same final state.
An approximation of the
MLM matching~\cite{Mangano} (referred to as
{\it ad hoc} matching) is used in the present
analysis, where the
matching is performed between matrix element partons and reconstructed
jets.
The \wplus MC samples are used in the analysis according to
the number of heavy flavor ($c$ or $b$) jets in the final state, classified
as follows:
$W+$~light denotes events without $c$ or $b$ jets;
$Wc$ denotes events with one $c$ jet due to single $c$ production;
$W(c\bar c)$ denotes events with one $c$ jet due to double $c$ production
where two $c$ quarks are merged in one jet or one of the $c$ jets is
outside of the acceptance region;
$Wc\bar c$ denotes events with two $c$ jets;
$W(b\bar b)$ denotes events with one $b$ jet due to double $b$ production
where two $b$ quarks are merged in one jet or one of the $b$ jets is
outside of the acceptance region (single $b$ production is highly
suppressed and neglected); and
$Wb\bar b$ denotes events with two $b$ jets.
Events are kept in the sample if the number of reconstructed jets equals the
      number of matrix element partons, where $(c\bar c)$ and $(b\bar
      b)$ are treated as one parton.
As the fourth jet multiplicity bin is treated inclusively in the analysis,
all events with $\ge 4$ reconstructed jets are kept,
independently of the number of additional non-matched light jets.

\section{Handling of the Event Statistics uncertainties}
\label{sec:stat_xs}
The matrix method (see Sec.~\ref{sec:matrixmethod})
is used three times in this analysis:
to determine the number of $W$-like
and QCD multijet events in the preselected, the single, and the double tagged
samples. The number of observed events used by the matrix method
is subject to random fluctuations according to Poisson statistics and
contributes to the total statistical uncertainty on the cross section
measurement. To treat these uncertainties properly, each number of events
entering the matrix method is considered as a free parameter constrained
to its observed value. This appendix details the treatment
of statistical uncertainties arising from the number of events
observed in data in the extraction of the cross section.

For the preselected samples, the matrix method
gives the number of $W$-like $N^{\text{sig}}$ and QCD multijet $N^{\text{QCD}}$
events in the tight preselected sample as
\[
N^{\text{sig}}_{t} = \varepsilon_{\text{sig}} \frac{N_{t} - \varepsilon_{\text{QCD}}
N_{\ell}}{\varepsilon_{\text{sig}} - \varepsilon_{\text{QCD}}},
\]
\[
N^{\text{QCD}}_{t} = \varepsilon_{\text{QCD}} \frac{\varepsilon_{\text{sig}} N_{\ell}
- N_{t}}{\varepsilon_{\text{sig}} - \varepsilon_{\text{QCD}}}.
\]
The true values $N_{\ell}$ and $N_{t}$
are not known, and are left floating
in the cross section calculation but constrained to their
measured values $\tilde{N}_{\ell}$ and $\tilde{N}_{t}$
using Poisson statistics.

It is necessary to take into account
that $N_{\ell}$
and $N_{t}$ are not independent variables.
To do so, the matrix method equations are expressed
in terms of $N_{t}$ and $N_{\ell-t}$, the latter representing
the number of events that are loose but not tight.
The equations become
\[
N^{\text{sig}}_{t}  = \varepsilon_{\text{sig}} \frac{N_{t} - \varepsilon_{\text{QCD}}
(N_{t}+N_{\ell-t})}{\varepsilon_{\text{sig}} - \varepsilon_{\text{QCD}}},
\]
\[
N^{\text{QCD}}_{t}  =  \varepsilon_{\text{QCD}} \frac{\varepsilon_{\text{sig}} (N_{t}
+ N_{\ell-t} ) - N_{t}}{\varepsilon_{\text{sig}} - \varepsilon_{\text{QCD}}}.
\]
Here, $N_{t}$ and
$N_{\ell-t}$ are constrained respectively to the observed number of
tight events, and to the observed number
of loose-but-not-tight events by adding the
following factor to the likelihood function
\[
{\mathcal P}(\tilde{N}_{t} ; N_{t})
  \times {\mathcal P}(\tilde{N}_{\ell-t} ; N_{\ell-t}),
\]
which represents the probability to observe
$\tilde{N}_{t}$ and $\tilde{N}_{\ell-t}$
given their true values $N_{t}$ and  $N_{\ell-t}$.

This procedure can be repeated for the single and double tagged samples to
predict the number of QCD multijet events as
\[
N_{\text{QCD}}^{\text{1tag}}    =  \varepsilon_{\text{QCD}} \frac{\varepsilon_{\text{sig}}
(N^{\text{1tag}}_{\ell-t} + N^{\text{1tag}}_{t}) - N^{\text{1tag}}_{t}}{\varepsilon_{\text{sig}} -
\varepsilon_{\text{QCD}}},
\]
\[
N_{\text{QCD}}^{\text{2tag}}    =  \varepsilon_{\text{QCD}} \frac{\varepsilon_{\text{sig}}
(N^{\text{2tag}}_{\ell-t} + N^{\text{2tag}}_{t}) - N^{\text{2tag}}_{t}}
{\varepsilon_{\text{sig}} - \varepsilon_{\text{QCD}}}.
\]
Note that the number of tight events with one tag $N^{1\text{tag}}_{t}$ and the number of tight events with two tags $N^{2\text{tag}}_{t}$ correspond to
$N^{\text{obs}}_{\gamma}$ in Eq.~\ref{eq:poisson1} in Sec.~\ref{sec:xs}.
Therefore,
$N^{1\text{tag}}_{t}$ and $N^{2\text{tag}}_{t}$ are already constrained to their
observed values and only one additional constraint
for the number of events in the
$\rm{loose}-{\rm tight}$ sample with one and two tags is needed:
\[
{\mathcal P}(\tilde{N}^{\text{1tag}}_{\ell-t} ; N^{\text{1tag}}_{\ell-t})
 \times  {\mathcal P}(\tilde{N}^{\text{2tag}}_{\ell-t} ; N^{\text{2tag}}_{\ell-t}),
 \]
which represents the probability to observe
$\tilde{N}^{1\text{tag}}_{\ell-t}$ and $\tilde{N}^{2\text{tag}}_{\ell-t}$ given their
true values $N^{1\text{tag}}_{\ell-t}$ and $N^{2\text{tag}}_{\ell-t}$.

Both the tight and the
$\rm{loose}-{\rm tight}$ sample can be separated into events with
zero, one, or two tags. Let $N^{\text{0tag}}_{t}$ and  $N^{\text{0tag}}_{\ell-t}$
represent the number of events with zero tags in the tight and the
$\rm{loose}-{\rm tight}$ sample, respectively. During the maximization
process, $N^{\text{0tag}}_{t}$ and  $N^{\text{0tag}}_{\ell-t}$ are
two free parameters that are constrained to their observed values
with Poisson probabilities
\[
{\mathcal P}(\tilde{N}^{\text{0tag}}_{\ell-t} ; N^{\text{0tag}}_{\ell-t})
 \times  {\mathcal P}(\tilde{N}^{\text{0tag}}_{t} ;
 N^{\text{0tag}}_{t}).
\]
In addition, the number of predicted
tagged events can be expressed in terms of the number of expected tagged events
originating from \ttbar, QCD multijet, $W$+jets, and other small electroweak
backgrounds, for one and two tags, respectively:
\[
  N^{1\text{tag}}_{t}  =  P^{1\text{tag}}_{t\bar t} N_{t\bar t} +
  N^{1\text{tag}}_{\text{QCD}} + P^{1\text{tag}}_{W} N_{W}
  + P^{1\text{tag}}_{\text{MC~bkg}} N_{\text{MC~bkg}},
\]
\[
  N^{2\text{tag}}_{t}  =  P^{2\text{tag}}_{t\bar t} N_{t\bar t} + N^{2\text{tag}}_{\text{QCD}} + P^{2\text{tag}}_{W} N_{W}
  + P^{2\text{tag}}_{\text{MC~bkg}} N_{\text{MC~bkg}}.
\]
The contribution from the small electroweak backgrounds
(diboson, single top, and $Z \rightarrow \tau\tau$ production) is labeled MC~bkg
to indicate that its normalization before tagging is obtained from MC.
$P^{1\text{tag}}_{\text{process}}$ and $P^{2\text{tag}}_{\text{process}}$ are the average event
tagging probability for a certain process,
for single and double tags, respectively.

The number of $W$+jets events in the preselected sample is given by
\[
 N_{W} = N_{t}^{\text{sig}} - N_{t\bar t} - N_{\text{MC~bkg}}.
\]
Substituting this expression for $N_{W}$
into the equations for $N^{1\text{tag}}_{t}$ and $N^{2\text{tag}}_{t}$
above allows us to express the latter quantities as a function
of the tagging
probabilities; signal and background efficiencies used in
the matrix method; MC prediction for the small electroweak processes;
and the floating parameters $N^{0\text{tag}}_{t}$,  $N^{0\text{tag}}_{\ell-t}$,
$N^{1\text{tag}}_{\ell-t}$, and $N^{2\text{tag}}_{\ell-t}$.
$N^{1\text{tag}}_{t}$ and $N^{2\text{tag}}_{t}$ are constrained to their
observed values using Poisson statistics
\[
      {\mathcal P}(\tilde{N}^{1\text{tag}}_{t} ; N^{1\text{tag}}_{t})
  \times
  {\mathcal P}(\tilde{N}^{2\text{tag}}_{t} ; N^{2\text{tag}}_{t}).
\]

The resulting likelihood is given by ${\mathcal L}_1$ below.
The index $i$ indicates the product over the channels
$e$+3 jets, $e$+4 jets, $\mu$+3 jet, and $\mu$+4 jets, respectively.

\begin{eqnarray*}
  \label{eq:poisson6}
  \mathrm{{\mathcal L}}_1 =
  \prod_i & \{ &
  {\mathcal P}(\tilde{N}^{0\text{tag}}_{t} ; N^{0\text{tag}}_{t})
  \times
  {\mathcal P}(\tilde{N}^{1\text{tag}}_{t} ; N^{1\text{tag}}_{t}) \\ \nonumber
  & \times &
  {\mathcal P}(\tilde{N}^{2\text{tag}}_{t} ; N^{2\text{tag}}_{t})
  \times
  {\mathcal P}(\tilde{N}^{0\text{tag}}_{\ell-t} ; N^{0\text{tag}}_{\ell-t}) \\ \nonumber
  & \times &
  {\mathcal P}(\tilde{N}^{1\text{tag}}_{\ell-t} ; N^{1\text{tag}}_{\ell-t})
  \times
  {\mathcal P}(\tilde{N}^{2\text{tag}}_{\ell-t} ; N^{2\text{tag}}_{\ell-t})
  \} \;. \nonumber
\end{eqnarray*}


\end{document}